\documentclass[modern, a4paper, times]{aastex62}
\usepackage[english]{babel}
\usepackage{amsmath}
\usepackage{graphicx}
\usepackage{natbib}
\usepackage{rotating}

\submitjournal{ApJ}

\shorttitle{Ionization degree of the interstellar matter around the Sun}

\begin{document}

\title{Interstellar neutral helium in the heliosphere from IBEX observations. VI. The He$^+$ density and the ionization state in the Very Local Interstellar Matter}

\correspondingauthor{M. Bzowski}
\email{bzowski@cbk.waw.pl} 

\author[0000-0003-3957-2359]{M. Bzowski}
\affil{Space Research Centre PAS (CBK PAN), Bartycka 18a, 00-716 Warsaw, Poland}

\author[0000-0002-4441-5377]{A. Czechowski}
\affil{Space Research Centre PAS (CBK PAN), Bartycka 18a, 00-716 Warsaw, Poland}

\author{P.C. Frisch}
\affil{University of Chicago}

\author[0000-0003-4101-7901]{S.A. Fuselier}
\affil{Southwest Research Institute, San Antonio, TX}
\affil{University of Texas at San Antonio, San Antonio, TX}

\author[0000-0003-2425-3793]{A. Galli}
\affil{University of Bern, Bern, Switzerland}

\author[0000-0003-3951-0043]6{J. Grygorczuk}
\affil{Space Research Centre PAS (CBK PAN), Bartycka 18a, 00-716 Warsaw, Poland}

\author[0000-0001-7867-3633]{J. Heerikhuisen}
\affil{University of Alabama in Huntsville, Huntsville, AL}

\author[0000-0002-5204-9645]{M.A. Kubiak}
\affil{Space Research Centre PAS (CBK PAN), Bartycka 18a, 00-716 Warsaw, Poland}

\author{H. Kucharek}
\affil{University of New Hampshire, Durham, NH}

\author[0000-0002-9745-3502]{D.J. McComas}
\affil{Department of Astrophysical Sciences, Princeton University, Princeton, NJ}

\author[0000-0002-2745-6978]{E. M{\"o}bius}
\affil{University of New Hampshire, Durham, NH}

\author[0000-0002-3737-9283]{N.A. Schwadron}
\affil{University of New Hampshire, Durham, NH}

\author[0000-0002-7597-6935]{J. Slavin}
\affil{Harvard-Smithsonian Center for Astrophysics, Cambridge, MA}

\author[0000-0002-4173-3601]{J.M. Sok{\'o}{\l}}
\affil{Space Research Centre PAS (CBK PAN), Bartycka 18a, 00-716 Warsaw, Poland}

\author[ 0000-0002-9033-0809]{P. Swaczyna}
\affil{Space Research Centre PAS (CBK PAN), Bartycka 18a, 00-716 Warsaw, Poland}
\affil{Department of Astrophysical Sciences, Princeton University, Princeton, NJ}

\author[0000-0002-2603-1169]{P. Wurz}
\affil{University of Bern, Bern, Switzerland}

\author[0000-0001-7240-0618]{E.J. Zirnstein}
\affil{Department of Astrophysical Sciences, Princeton University, Princeton, NJ}

\begin{abstract}
Interstellar neutral gas atoms penetrate the heliopause and reach 1~au, where they are detected by  IBEX. The flow of neutral interstellar helium through the perturbed interstellar plasma in the outer heliosheath (OHS) results in creation of the secondary population of interstellar He atoms, the so-called Warm Breeze, due to charge exchange with perturbed ions. The secondary population brings the imprint of the OHS conditions to the IBEX-Lo instrument. Based on a global simulation of the heliosphere with measurement-based parameters and detailed kinetic simulation of the filtration of He in the OHS, we find the number density of interstellar He$^+$ population at $(8.98\pm 0.12)\times 10^{-3}$~cm$^{-3}$. With this, we obtain the absolute density of interstellar H$^+$ $5.4\times 10^{-2}$~cm$^{-3}$ and electrons $6.3\times 10^{-2}$~cm$^{-3}$, and ionization degrees of H 0.26 and He 0.37. The results agree with estimates of the Very Local Interstellar Matter parameters obtained from fitting the observed spectra of diffuse interstellar EUV and soft X-Ray background.
\end{abstract}

\section{Introduction}
\label{sec:intro}
\noindent
The Sun is traversing a cloud of dilute ($\sim 0.25$~nucleons~cm$^{-3}$), warm ($\sim 7500$ K), partly ionized, magnetized ($\sim 3 \mu$G) interstellar matter that is an element of a larger cloud structure. This complex cloud is described in the literature either as a set of relatively small individual interstellar clouds \citep[see, e.g., ][]{redfield_linsky:08a} or a more sizable and complex structure with large-scale internal motions \citep{frisch_etal:02a, gry_jenkins:14a}. It is located inside the Local Bubble (LB) of low-density ($\sim 10^{-3}$~nuc~cm$^{-3}$), hot ($\sim 1$~MK), fully ionized plasma, which most likely formed by overlapping superbubbles from recent nearby supernovae explosions \citep{frisch_etal:11a}. The portion of interstellar matter that fills the immediate neighborhood of the Sun and thus is accessible to local observations will be referred to here as the Very Local Interstellar Matter (VLISM). The physical state of the matter within the VLISM is determined by collisional, charge exchange, and recombination processes, radiative cooling, ionization and heating by EUV radiation from nearby stars, the LB, and with possible contribution from a nearby conductive interface between interstellar clouds and the LB \citep{slavin_frisch:08a}. Because of gradual absorption of the ionizing radiation inside the VLISM, the densities of the VLISM components, and consequently the local spectrum of the EUV radiation, vary with location in space. 

The physical state, elemental composition, and ionization state of the VLISM are studied by fitting parameters of models to observed profiles of interstellar absorption lines and the spectrum of the soft diffuse X-ray \citep{mccammon_etal:83a, bloch_etal:86a} and EUV \citep{vallerga_etal:04a} sky background. These are line-of-sight integrated observations of an inhomogeneous medium. Hence, investigating the physical state of the VLISM requires making parametric studies that consider cloud opacity \citep{slavin_frisch:08a}. Among the quantities challenging to obtain, but important for the VLISM physics, are absolute densities of the two main constituents of interstellar matter, hydrogen and helium. Their ionized fractions are especially important, because H$^+$ does not produce absorption lines, and available observations of He$^+$ lines are limited \citep{wolff_etal:99}. Even more challenging is investigating the magnetic field: this has been done on the global VLISM-scale by analysis of the polarization of starlight by field-aligned grains of interstellar dust \citep{frisch_etal:15c}.

Some of the VLISM parameters can be retrieved from direct sampling of ISN atoms and their daughter products in the solar wind, i.e., pickup ions (PUIs), and thus do not require global modeling of the heliosphere. They enable obtaining the density of ISN H and He, their temperature, and the direction and speed of Sun's motion through the VLISM.

Additional insight into the VLISM is obtained from comparing observations with predictions from global heliospheric modeling. The interaction of the solar wind and interstellar plasmas is modeled within a magnetohydrodynamic framework, coupled with a kinetic description of the plasma--neutral gas interaction \citep{baranov_malama:93,pogorelov_etal:09a, heerikhuisen_etal:10a}. Results of this modeling provide predictions for various observable quantities, including the heliopause distance, plasma density, and magnetic field at the Voyager spacecraft, as well as the size and location of the IBEX Ribbon.

The flux of ISN He atoms observed by the IBEX-Lo instrument \citep{fuselier_etal:09b, mobius_etal:09a} onboard the Interstellar Boundary Explorer (IBEX) mission \citep{mccomas_etal:09a} was resolved into the primary \citep{mobius_etal:09b} and secondary \citep{kubiak_etal:14a} ISN populations. The secondary population is expected from heliospheric models \citep{baranov_malama:93} as a result of charge exchange between the plasma flowing by the heliopause and the unperturbed neutral gas in the OHS, i.e., the region ahead of the heliosphere where interstellar matter adapts to flow past the heliopause. For He, the dominant reaction is charge exchange between ISN He atoms and He$^+$ ions \citep{bzowski_etal:12a}. \citet{bzowski_etal:17a} (Bz2017) simulated the ISN He signal observed by IBEX accounting for this charge exchange process in the OHS. They showed that the simulated signal is sensitive to the conditions in the OHS and VLISM, among others to the plasma density and magnetic field vector. 

Here, we constrain the absolute densities of the H$^+$ and He$^+$ components of the VLISM by fitting the ISN He signal observed by IBEX using the model from Bz2017 in conjunction with the global model of the heliosphere from \citet{zirnstein_etal:16b} (Zir2016). Observations and data selections are presented in Section~\ref{sec:datasel}. Simulating the IBEX signal is described in Section~\ref{sec:signalsimul}, and the choice of parameter values used in the global heliospheric modeling performed for this study is discussed in Section~\ref{sec:simulparams} and presented in Table~\ref{tab:simulParams}. The method used to obtain the He$^+$ densities is discussed in Section~\ref{sec:paramfitting}, and the VLISM parameters obtained are shown in Section~\ref{sec:VLISMParamDeriv} and Table~2. 
Discussion of the conclusions on the physical state of the VLISM concludes the paper.

\section{Observations and data selection}
\label{sec:datasel}
\begin{figure}
 \plotone{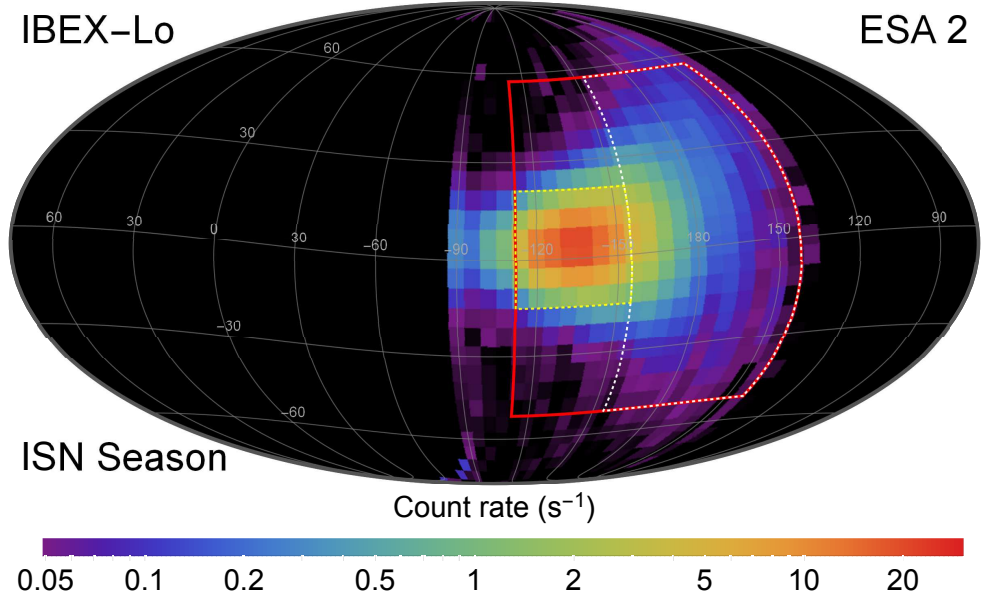}
 \caption{Sky map of the observed IBEX-Lo count rates from enerfy step ESA 2 used for analysis, averaged over the ISN observation seasons 2010--2014 (color-coded). The contours mark the data ranges used by \citet{bzowski_etal:15a} to derive the ISN He parameters (yellow), by Ku2016 to derive the parameters of the secondary population of ISN He in the two-Maxwellian approximation (white), and in this work to derive interstellar He$^+$ density in the VLISM (red).}
 \label{fig:datamap}
 \end{figure}

The organization of the ISN gas sampling by IBEX-Lo was presented in detail by \citet{mobius_etal:09a, mobius_etal:12a, mobius_etal:15b}, and here we only recall the most important aspects. IBEX is a spin-stabilized spacecraft orbiting the Earth in a highly elongated orbit with an apogee up to $\sim 50$ Earth radii. The spin axis of the spacecraft never points farther than $\sim 7\degr$ from the Sun. The optical axis of the IBEX-Lo instrument is perpendicular to the spin axis. The atom detection events are accumulated in time so that they correspond to 6\degr~bins of the spacecraft spin angle. For this study, we used the counting rate registered in energy step 2 (ESA 2) of IBEX-Lo in each of the spin angle bins, averaged over the time intervals regarded as free from magnetospheric and solar wind contamination \citep{galli_etal:15a, galli_etal:16a}. IBEX ISN observations are carried out between November and March each year.

The data we use here are from the same observation seasons as those used by \citet[][Ku2016]{kubiak_etal:16a}  to facilitate comparisons. Following these authors, we adopted counts from IBEX-Lo ESA 2, spin angle range from 216\degr~to 318\degr, collected during the observation seasons 2010--2014. Ku2016 restricted the data to the IBEX orbits where the secondary population (Warm Breeze) signal dominated (which corresponds to the range of ecliptic longitudes of IBEX spin axis $235\degr$ to $295\degr$). We use the yearly intervals corresponding both to the Warm Breeze and the primary ISN He signal, selecting the data based on an improved list of ``good observation times'' after \citet{galli_etal:16a}. We took orbits with both the Warm Breeze and the primary ISN signal because in our signal synthesis method we simulate both populations simultaneously, without differentiating between them. We cut off the portion of the data taken for spin axis longitudes larger than $\sim 335\degr$ to avoid a contribution to the signal from ISN H \citep{swaczyna_etal:18a, galli_etal:19a}. As a consequence, we have significantly more data points (1422) than used by \citet{bzowski_etal:15a} and Ku2016  (972). A comparison of the data range used by these and in this study is shown in Figure~\ref{fig:datamap}. The data we used are available in IBEX data release 12 at http://ibex.swri.edu/researchers/publicdata.shtml. 

\section{IBEX signal simulation}
\label{sec:signalsimul}
\noindent
The signal simulation process is similar to that used by Bz2017. First, a global model of the heliosphere is run to provide the heliopause location and the plasma flow in the OHS. Subsequently, the observed signal is synthesized by integrating the contributions to the observed flux from individual He atoms. The statistical weights for these atoms are obtained from solutions of the production and loss balance equation. This equation is solved along the atom trajectories from the unperturbed interstellar medium through the outer heliosheath down to IBEX at 1~au. 

We have verified that the region in the sky where atoms reaching IBEX-Lo penetrate the heliopause is oval-like, centered at the direction of inflow of the Warm Breeze (the secondary ISN He) found by Ku2016, and extends approximately $\pm 60\degr$ in longitude and latitude (see Figure~\ref{fig:NDP} and Figure 7 in Bz2017). This is the region in the OHS ballistically connected to the IBEX-Lo instrument. This implies that predictions of any heliospheric model outside this region do not affect our results. 

\subsection{Boundary conditions for ISN He}
\label{sec:boundCond}
\noindent
In the inertial reference system co-moving with the Sun, interstellar matter is inflowing on the heliosphere with the velocity $\vec{u}_\text{VLISM}$. The distribution function of ISN He in the unperturbed VLISM far ahead of the heliosphere is assumed to be the Maxwell-Boltzmann distribution:
\begin{equation}
f_{\text{He}_{\text{VLISM}}}(\vec{v}) = n_{\text{He}_\text{VLISM}}\left(\frac{\,m_\text{He}}{2 \pi k_\text{B} T_\text{VLISM}}\right)^{\frac{3}{2}}\,\exp\!\left[ -\frac{m_\text{He}(\vec{v}-\vec{u}_{\text{VLISM}})^2}{2 k_\text{B} T_\text{VLISM}}\right],
\label{eq:fMBVLISM}
\end{equation}
where $\vec{v}$ is the velocity vector of an individual atom, the density of ISN He is $n_{\text{He,VLISM}}$, and $(2k_\text{B}T_{\text{VLISM}}/m_{\text{He}})^{1/2} = u_{\text{T,VLISM}}$ is the thermal speed of He in the VLISM for the temperature $T_{\text{VLISM}}$. This definition corresponds to Equation~6 in Bz2017; note that Bz2017 have a mistake in the denominator of the scaling factor in front of the exponent function. 

\subsection{He$^+$ ions in the OHS}
\label{sec:heplOHS}
\noindent
The properties of the plasma in the OHS are adopted from the simulation of the interaction of the magnetized VLISM matter with the solar wind carried out using the Huntsville global MHD model of the heliosphere \citep{heerikhuisen_pogorelov:10a} with the simulation parameters identical to those found by Zir2016 (see Table~\ref{tab:simulParams}). This model uses proton plasma in the OHS and neglects all heavy ion components, including He$^+$. In reality, He$^+$ ions contribute significantly to the ram pressure because they have a $\sim 4$ times larger mass and a typical He-to-H number ratio in the astrophysical plasma is of the order of 10\%. Here, we interpret the plasma flow obtained in the model as a sum of a co-moving mixture of H$^+$ and He$^+$ ions. Moreover, the ratio of these two components is assumed constant throughout the outer heliosheath. Consequently, the density of He$^+$ ions in the OHS is 
\begin{equation}
n_{\text{He}^+_\text{OHS}}(\vec{r})=n_{\text{He}^+_\text{VLISM}}\,\frac{\rho_{\text{pl}_\text{OHS}}(\vec{r})}{\rho_{\text{pl}_\text{VLISM}}},
\label{eq:nheplus}
\end{equation}
where $\rho_{\text{pl}_\text{OHS}}(\vec{r})$ is the assumed plasma density in the OHS from the global model,  $\rho_{\text{pl}_\text{VLISM}}$ is the plasma density in the unperturbed VLISM, and $n_{\text{He}^+_\text{VLISM}}$ is the sought He$^+$ density in the unperturbed VLISM. Further, we assume for H$^+$ and He$^+$ ions the same plasma flow $\vec{u}_{\text{OHS}}(\vec{r})$ and temperature $T_{\text{OHS}}(\vec{r})$, resulting from the global model. The plasma parameters in the OHS are interpolated between the grid nodes.

The distribution function of  He$^+$ ions in the OHS $f_{\text{He}^+_\text{OHS}}$ is assumed to be Maxwell-Boltzmann
\begin{equation}
f_{\text{He}^+_\text{OHS}}\left(\vec{r}, \vec{v}\right) = n_{\text{He}^+_\text{OHS}}(\vec{r})  \left(\frac{m_\text{He}}{2 \pi k_\text{B} T_\text{OHS}(\vec{r})}\right)^{\frac{3}{2}}\,\exp\!\left[-\frac{m_\text{He}\left(\vec{v} - \vec{u}_\text{OHS}(\vec{r})\right)^2}{2k_\text{B}T_\text{OHS}(\vec{r})} \right],
\label{eq:plasmaDiFu}
\end{equation}
where $m_\text{He}$ is the mass of He$^+$ ion. 

\subsection{Statistical weights of ISN He}
\label{sec:statWeights}
\noindent
The calculation of the statistical weights presented in this section was done identically as in Bz2017. Here, we present a description of this process that, in our opinion, better highlights the suitability of this method to obtain the absolute density of ISN He$^+$.

The quantity used to calculate the signal is the statistical weight $\omega_{\text{loc}}=\omega(\vec{r}_{\text{loc}},\vec{v}_{\text{loc}})$ at the location $\vec{r}_{\text{loc}}$ where an atom moving with the velocity $ \vec{v}_{\text{loc}}$ is detected. This quantity is obtained from numerical solution of the equation of production and loss balance (\ref{eq:prodlossBalance}) for He atoms in the trajectory $s$ defined by the state vector of the atom at the detector $\vec{q}_\text{stat} = (\vec{r}_{\text{loc}}, \vec{v}_{\text{loc}})$. Since the atom motion is purely Keplerian (hyperbolic trajectory), the atom orbit $s$ is uniquely determined by $\vec{q}_\text{stat}$. Therefore, any other state vector in this orbit $\vec{q}_s(t)$ can be obtained at any time. The relation between time and true anomaly (i.e., the heliocentric angle between the perihelion point of the orbit and the actual location of the atom in the orbit) is given by the hyperbolic Kepler equation. 

The production and loss balance equation is defined as follows:
\begin{equation}
\frac{d\omega(t)}{d t} = \beta_{\text{pr}}(\vec{r}_s(t),\vec{v}_s(t))\ \ f_{\text{He}^+_\text{OHS}}(\vec{r}_s(t),\vec{v}_s(t)) - \beta_{\text{loss}}(\vec{r}_s(t),\vec{v}_s(t))\ \ \omega(t),
\label{eq:prodlossBalance}
\end{equation}
where $\beta_{\text{pr}}\ f_{\text{He}^+_\text{OHS}}$ represents the production of new neutral He atoms from He$^+$ ions in the OHS, and $\beta_{\text{loss}}\ \omega$ the losses of neutral He atoms due to charge exchange with He$^+$ ions. Here, the distribution function of He$^+$ ions is a separate factor from the production rate $\beta_\text{pr}$, which were multiplied together in Equation~11 in Bz2017. The production and loss balance Equation \ref{eq:prodlossBalance} is solved for each considered state vector $\vec{q}_{s}$.

The production rate $\beta_\text{pr}$ in a given location within the OHS along the trajectory $s$ is determined by the rate of the resonant charge exchange reaction between ambient He atoms and He$^+$ ions:
\begin{eqnarray}
\beta_\text{pr}(\vec{r},\vec{v})&=&n_{\text{He}_\text{VLISM}}\,u_\text{rel}^\text{pr}\,\sigma_\text{cx}(u_\text{rel}^\text{pr}),\\
\label{eq:proRaeteDef}
u_\text{rel}^\text{pr}&=&u_\text{rel}(|\vec{v}-\vec{u}_{\text{VLISM}}|, u_{\text{T}_\text{VLISM}}),
\label{eq:urelpr}
\end{eqnarray}
where $\sigma_\text{cx}$ is the reaction cross section, and the mean relative speed $u_\text{rel}^\text{pr}$ is between an ion traveling at $\vec{v}$ and a Maxwell-Boltzmann population of the ambient He atoms with the temperature $T_\text{VLISM}$, bulk velocity $\vec{u}_\text{VLISM}$, and density $n_\text{He,VLISM}$, which are assumed constant throughout the OHS in this equation. The reaction involves no momentum exchange between the collision partners. The mean relative speed is calculated as
\begin{equation}
u_\text{rel}(\Delta u, u_\text{T}) =  \left[ \left( \frac{\Delta u}{u_\text{T}} + \frac{u_\text{T}}{2\Delta u} \right)\text{erf}\left( \frac{\Delta u}{u_\text{T}} \right)+\frac{\exp(-\Delta u^2/u_\text{T}^2 )}{\sqrt{\pi}}\right]u_\text{T}.
\label{eq:relvel}
\end{equation}
This formula appropriately weighs the velocity of the impactor particle relative to the centroid of the target particles and the thermal speed of the target particles. It has been used in a number of papers, including the original paper by \citet{fahr_mueller:67} and, importantly, by \citet{heerikhuisen_etal:15a} and Zir2016 in their modeling of the heliosphere. 

The loss rate $\beta_\text{loss}$ of He atoms at a location in the trajectory $s$ is proportional to the sum of photoionization and charge exchange ionization rates. The magnitude of the photoionization rate in the OHS is assumed to be constant in time and equal to $\beta_{\text{ph},0} (r_\text{E}/r)^2$, with $\beta_{\text{ph},0}$ equal to the helium photoionization rate at $r_\text{E}=1$~au, averaged over the solar cycle (approximately $10^{-7}$~s$^{-1}$). The losses via charge exchange are due to collisions between He atoms and He$^+$ ions from the ambient OHS plasma. Consequently, the loss term is given by:
\begin{eqnarray}
\beta_\text{loss}(\vec{r},\vec{v}) &=& \beta_\text{ph,0}\left(\frac{r_\text{E}}{r} \right)^2 + n_{\text{He}^+_\text{OHS}}(\vec{r})u_\text{rel}^\text{loss} \sigma_\text{cx}(u_\text{rel}^\text{loss}),
\label{eq:lossrateDef}\\
u_\text{rel}^\text{loss}&=&u_\text{rel}(|\vec{v}-\vec{u}_{\text{OHS}}(\vec{r})|, u_{\text{T,OHS}}(\vec{r})),
\label{eq:urelloss}
\end{eqnarray}
where $u_{\text{T}_\text{OHS}}(\vec{r})=(2k_\text{B}T_{\text{OHS}}(\vec{r})/m_{\text{He}})^{1/2}$.  Note that for calculation of the loss rates, the density, flow velocity, and temperature of the ambient He$^+$ ion population are obtained from the global model of the heliosphere.

The solution of the production and loss balance equation starts at a limiting distance, adopted as LD = 1000~au from the Sun. First, based on the orbital parameters obtained from the state vector $\vec{q}_{\text{stat}}$ at the detector location, the velocity $\vec{v}_s(t_{\text{LD}})$ at the limiting distance of the calculation is determined. Beyond this distance, all interstellar populations are assumed to be homogeneous in space and in collisional equilibrium. With that, the initial condition for Equation~\ref{eq:prodlossBalance} $\omega_0$ is defined as:
\begin{equation}
\omega_0 = f_{\text{He}_\text{VLISM}}(\vec{v}_s(t_{\text{LD}})), 
\label{eq:omegaVLISM}
\end{equation}
with $f_{\text{He}_\text{VLISM}}$ defined in Equation~\ref{eq:fMBVLISM}. Note that, a priori, any reasonable form of $f_{\text{He}_\text{VLISM}}$ can be adopted, including, e.g., a kappa function, as that used by \citet{sokol_etal:15a}, but if so, then consequently both in the global simulation and in the calculation of the statistical weights. Moreover, since all terms in Equation~\ref{eq:prodlossBalance} are proportional to the density of ISN He in the VLISM ($\omega,\, \beta_\text{pr} \propto n_{\text{He}_\text{VLISM}}$), this density cannot be obtained from IBEX observations without relying on the absolute instrument calibration. However, only the loss rate and the distribution function of He$^+$ in the OHS are proportional to the density of He$^+$ ions ($\beta_\text{loss},\, f_{\text{He}^+_\text{OHS}} \propto n_{\text{He}^+_\text{VLISM}}$). Therefore, the absolute density of He$^+$ can be retrieved from comparison of statistical weights $\omega$ obtained for the atom orbits with different statistical weights $\omega_0$ at the limiting distance LD. The result does not depend on the magnitude of the adopted density of ISN He in the unperturbed VLISM.

For the He$^+$ + He charge exchange cross section, we used the formula from \citet{barnett_etal:90}, adapted for low collision speeds of a few km~s$^{-1}$ (Equation~10 in Bz2017). We neglect the charge exchange between the He and H populations (neutral and ionized) because the cross sections for these reactions are $\sim 250$ times lower than those for the He$^+$ + He $\rightarrow$ He + He$^+$ reaction for the low collision speeds ($\sim 5-50$~km~s$^{-1}$), characteristic for the charge exchange collisions in the OHS. 

With the initial condition defined, Equation~\ref{eq:prodlossBalance} is solved along the trajectory down to the detector. Solving this equation is done in two parts: first between the source region and the heliopause, i.e., within the OHS, and subsequently from the heliopause to the detector. 

For the solution of Equation~\ref{eq:prodlossBalance} within the OHS we assume a time-stationary situation. With this, Equation~\ref{eq:prodlossBalance} is transformed so that the integration goes over the true anomaly angle $\theta$, which becomes the independent variable: $dt = \frac{r^2}{L}\,d\theta$, where $L= |\vec{L}| = |\vec{r}_\text{loc} \times \vec{v}_{\text{loc}}|$ is the angular momentum per unit mass, constant over the atom orbit. Inside the heliopause, calculating the losses is carried out identically as in our previous papers presenting analyses of IBEX ISN He observations: \citet{bzowski_etal:12a, kubiak_etal:14a, bzowski_etal:15a, kubiak_etal:16a, swaczyna_etal:18a}. 

Ultimately, the solution returns the statistical weight $\omega_{\text{loc}}$ of a given atom at the detector, which can be regarded as the magnitude of the local distribution function of neutral He at the detector site for the atom state vector $\vec{q}_s$.

\subsection{Integration with IBEX-Lo response}
\label{sec:integrLoRespo}
\noindent
The calculation scheme of the IBEX-Lo signal due to ISN He takes into account all important details of the data taking procedure so that the simulated signal can be directly compared with the IBEX-Lo data product \citep{sokol_etal:15b}. In this paper, we precisely follow the approach described by these authors up to the point in the calculations where the magnitude of the local distribution function of ISN He for a given state vector of the atom at the detector is needed. There, the method is changed to that developed by Bz2017 and presented in detail in Sections~\ref{sec:boundCond}--\ref{sec:statWeights}.

\begin{figure}
\plotone{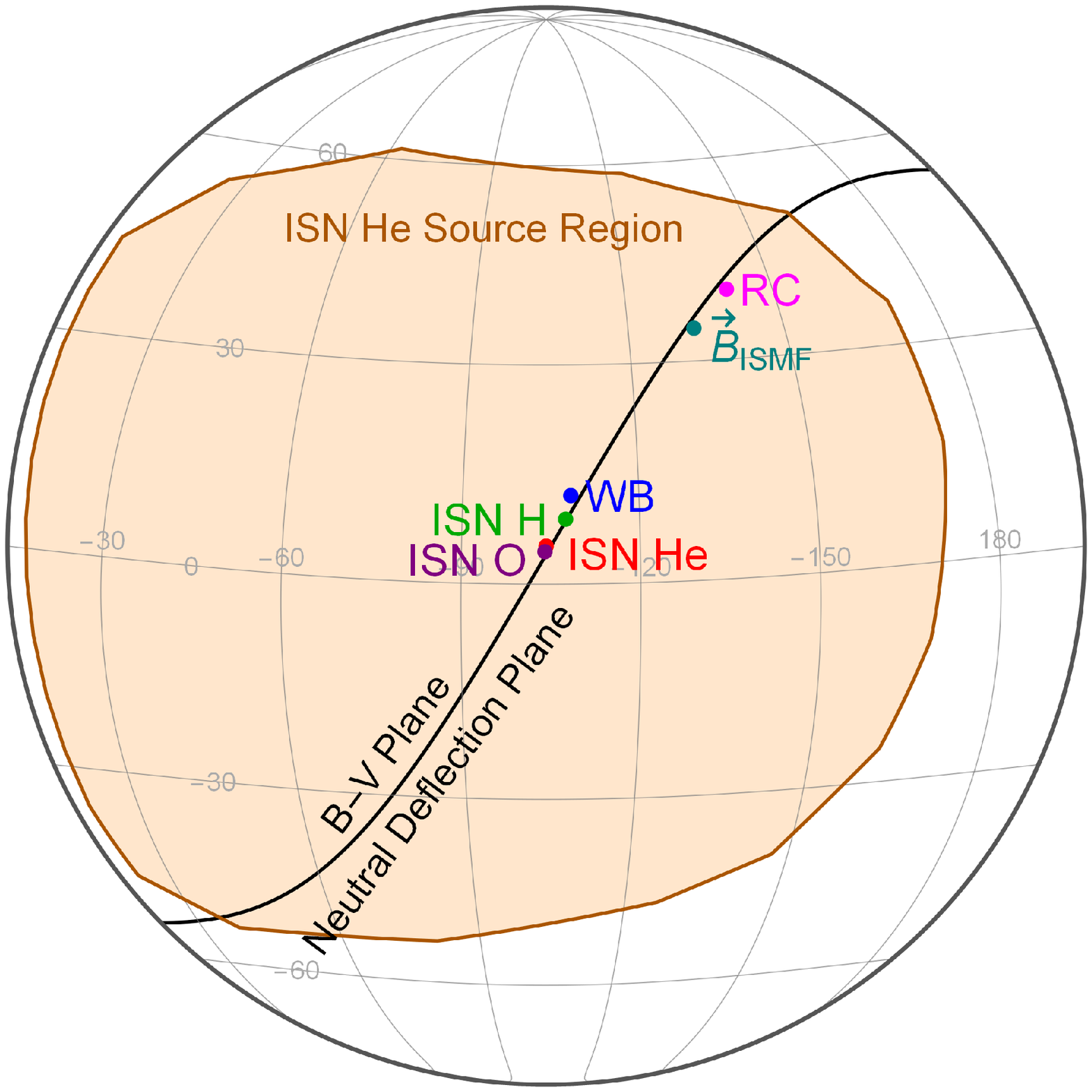}
\caption{The geometry of the Sun's motion through the VLISM, the ISMF vector ($\vec{B}_\text{ISMF}$), the Ribbon center (RC), and the secondary populations of ISN gas (``the Warm Breeze'', WB), projected on the sky. The direction of the Sun's motion through the VLISM and its uncertainty as obtained by \citet{bzowski_etal:15a}, is marked by ISN He. Also shown is the direction of ISN O \citep{schwadron_etal:16a}. The inflow direction of ISN H \citep{lallement_etal:05a} corresponds to a superposition of the primary and secondary populations. The average direction of RC is adopted following \citet{funsten_etal:13a}, and that of the unperturbed ISMF is adopted from Zir2016. The black line marks the Neutral Deflection Plane (NDP) obtained by Ku2016 from fitting the directions of ISN He, ISN H, and WB. The B-V plane used in the calculations includes the ISN He and $B_\text{ISMF}$ directions. The brown-shaded region is the projection on the sky of the locations at the heliopause where the He sampled by IBEX-Lo enter the heliosphere.}
\label{fig:NDP}
\end{figure}

\section{Parameter values for the simulations}
\label{sec:simulparams}
\noindent
In this section we present the parameters used in the two-tier simulations performed to obtain the He$^+$ density in the VLISM and argue that they are supported by a strong observational foundation. The adopted parameter values were used consistently in the global simulation of the heliosphere and in the simulations used to derive the He$^+$ density in the VLISM. They are collected in Table~\ref{tab:simulParams}.

\begin{deluxetable*}{rcl}
\tablecaption{Parameter values adopted in the calculations}
\tablehead{
\colhead{quantity}  &
\colhead{magnitude} &
\colhead{reference}
}
\startdata
VLISM plasma mass density $\rho_{\text{pl,VLISM}}$              & $0.09$~nuc~cm$^{-3}$\tablenotemark{a}                 & Zir2016 \\
VLISM neutral H density $n_\text{H,VLISM}$                      & $0.154$~cm~$^{-3}$\tablenotemark{a}\tablenotemark{b}  & Zir2016 \\
VLISM neutral He density $n_\text{He,VLISM}$                    & $0.0150$\tablenotemark{a}\tablenotemark{b}            & \citet{gloeckler_etal:04a} \\
VLISM temperature $T_{\text{VLISM}}$                            & $7500$~K\tablenotemark{a}\tablenotemark{c}            & \citet{mccomas_etal:15b}   \\
ISMF strength $B_{\text{ISMF}}$                                 & $2.93,\mu$G                                           & Zir2016 \\
ISMF direction $(\lambda_{\text{ISMF}}, \beta_{\text{ISMF}})$   & $(227.28\degr, 34.62\degr)$\tablenotemark{a}\tablenotemark{d} & Zir2016 \\
Sun's motion speed $u_{\text{VLISM}}$                           & $25.4$~km~s$^{-1}$                                    & \citet{mccomas_etal:15b}   \\
Sun's motion direction $\lambda_{\text{VLISM}},\beta_{\text{VLISM}}$ & $(255.7\degr, 5.1\degr)$\tablenotemark{a}\tablenotemark{d}\tablenotemark{e} & \citet{mccomas_etal:15b} \\
solar wind density at 1 au $n_{\text{SW}}$                      & $5.74$~cm$^{-3}$                                      & Zir2016 \\
solar wind temperature at 1 au $T_{\text{SW}}$                  & $5.1\times 10^4$~K                                    & Zir2016 \\
solar wind speed at 1 au $v_{\text{SW}}$                        & 450~km~s$^{-1}$                                       & Zir2016 \\
solar wind mgt field radial at 1 au $B_{\text{r,SW}}$           & $37.5\,\mu$G                                          & Zir2016 \\
\enddata
\tablenotetext{a}{Fixed value adopted in the simulations of ISN He in this paper.}
\tablenotetext{b}{Uncertainty for the determination of ISN H by \citet{bzowski_etal:09a} given at $0.16 \pm 0.04$~cm$^{-3}$; for the determination of ISN He by \citet{gloeckler_etal:04a} $\pm 0.0015$~cm$^{-3}$.}
\tablenotetext{c}{Affects the initial condition for production and loss balance equation~\ref{eq:prodlossBalance}. The ut ncertainty for the determination by \citet{bzowski_etal:15a} given at $7440\pm 260$~K but \citet{swaczyna_etal:18a} suggests $7700 \pm 230$ and \citet{schwadron_etal:15a} $8000\pm 1300$~K.}
\tablenotetext{d}{The directions of $\vec{u}_\text{VLISM}$ and $\vec{B}_\text{ISMF}$ determine the orientation of the B-V plane and consequently their uncertainties contribute to the uncertainty of the transformation from the reference system of the global heliosphere simulation to the ecliptic coordinates. The uncertainties for the direction of $\vec{B}_\text{ISMF}$ are $\pm 0.08\,\mu$G in strength and $(\pm 0.69\degr, \pm 0.45\degr)$ in (longitude, latitude). The strength of $B_\text{ISMF}$ does not directly affect our simulations of ISN He, only the global heliosphere model. }
\tablenotetext{e}{The uncertainties for longitude, latitude of $\vec{u}_\text{VLISM}$ are at least $(\pm 0.5\degr, 0.1\degr)$ and for speed $\pm 0.3$~km~s$^{-1}$ \citep{swaczyna_etal:18a}, but \citet{schwadron_etal:15a} gives uncertainties for (longitude, latitude) equal to $(\pm 1.4\degr, \pm 0.3\degr)$ and $\pm 1.1$~km~s$^{-1}$. 
The uncertainties of the VLISM velocity vector components and of $T_\text{VLISM}$ are strongly correlated with each other.}
\label{tab:simulParams}
\end{deluxetable*}

\subsection{VLISM parameters}
\label{sec:VLISMparams}
\subsubsection{Sun's velocity vector and ISN gas temperature}
\label{sec:SunVelocity}
\noindent
The vector of Sun's velocity relative to the VLISM and the VLISM temperature had been obtained from extensive analyses of direct-sampling observations of ISN He available from Ulysses \citep{witte:04, bzowski_etal:14a,wood_etal:15a} and IBEX \citep{bzowski_etal:15a, mccomas_etal:15a, mccomas_etal:15b, mobius_etal:15b, schwadron_etal:15a}. We adopted them following \citet{mccomas_etal:15b} to maintain homogeneity with Zir2016. The vector of inflow velocity of ISN gas on the heliosphere, used in Equation~\ref{eq:fMBVLISM}, is given by $\vec{u}_\text{VLISM} = -u_\text{VLISM}(\cos \lambda_\text{VLISM}\,\cos \beta_\text{VLISM},\sin \lambda_\text{VLISM}\,\cos \beta_\text{VLISM}, \sin \beta_\text{VLISM})$. These parameters are independent of heliosphere models since they rely on atom ballistics and ionization losses, obtained from measurements of relevant solar factors (Section~\ref{sec:solwindeuv}). 

Alternatively, the velocity of Sun's motion through the VLISM can be assessed from observations of the Doppler shifts of interstellar absorption lines visible in the spectra of nearby stars \citep{adams_frisch:77a}, using a method developed by \citep{crutcher:82a}. In this case, the result would be an average value over a distance of at least several parsecs. For our purpose, however, the value of the Sun's speed and VLISM temperature must be determined precisely at the Sun's location.  

\subsubsection{Mass density of interstellar plasma and number density of ISN H}
\label{sec:VLISMDensities}
\noindent
The plasma mass density value was chosen to provide a ram pressure needed, together with magnetic pressure, to obtain the heliopause distance in agreement with that found by Voyager 1: $\rho_{\text{pl,VLISM}} = 0.09$~nuc~cm$^{-3}$. This is the total mass density of the plasma, assumed to be composed of H$^+$ and He$^+$ ions. The expected contribution to the total mass density of the VLISM plasma from He$^{++}$ and heavy ions is negligible for the pressure balance \citep{slavin_frisch:08a}. This quantity is model-dependent, but the model used to derive it has been demonstrated to reproduce the target observable parameters (see details in Zir2016). This approach was qualitatively verified by measurement of the total electron density in the region of compressed plasma beyond the heliopause in situ by Voyager~1 \citep{gurnett_etal:13a, gurnett_etal:15a} at $\sim 0.09$~cm$^{-3}$, in qualitative agreement with predictions of the Huntsville heliosphere simulation for this region, reported by Zir2016. Also the intensity of the draped magnetic field in the Huntsville model was compliant with the Voyager measured values \citep{burlaga_ness:16a}.
 
The density of ISN H at the termination shock (TS) was determined using two independent methods. One of them is based on the magnitude of slowdown of solar wind inside the termination shock due to mass- and momentum loading from pickup of ISN H atoms ionized by charge exchange or photoionization \citep{isenberg:86, fahr_rucinski:99,lee_etal:09a}. This slowdown is proportional to the absolute density of ISN H at TS. Based on observations of Voyager 2 and Ulysses, it was measured by \citet{richardson:08a} to be $\sim 67$~km~s$^{-1}$. Using a one-dimensional MHD model of the slowdown, these authors determined the density of ISN H at TS equal to 0.09~atoms~cm$^{-3}$. 

The other method is based on the fact that the production rate of H$^+$ pickup ions at the boundary of the ISN H cavity is linearly dependent on the TS density of ISN H, but very weakly depends on all other parameters, including the solar resonant radiation pressure and the velocity, temperature, and ionization rate of ISN H. The PUI production rate was measured by SWICS on Ulysses \citep{gloeckler_geiss:01a}, and based on this measurement and modeling of the PUI production rate, the ISN H density at TS was determined by \citet{bzowski_etal:08a} to be $0.087 \pm 0.022$~cm$^{-3}$, in excellent agreement with the value obtained from the solar wind slowdown. \citet{kowalska-leszczynska_etal:18b} independently supported this conclusion using better models of radiation pressure and ionization losses. Based on the findings of \citet{richardson_etal:08a} and \citet{bzowski_etal:08a}, \citet{bzowski_etal:09a} suggested that the ISN H number density at TS is $0.09\pm 0.02$~cm$^{-3}$, and the number density in the unperturbed VLISM is $0.16 \pm 0.04$~atoms~cm$^{-3}$. The number density of ISN H at TS is independent of global heliospheric modeling. The transition from the TS density to the density in the unperturbed VLISM is done by heliospheric modeling \citep[e.g.,][]{izmodenov_etal:03a, izmodenov_etal:03b}, but the TS/VLISM ratio for $n_{\text{H}}$ only weakly depends on the details of these models \citep{bzowski_etal:08a}. In the global heliospheric simulation in our paper, we adopted the VLISM H density $n_{\text{H,VLISM}} = 0.154$~atoms~cm$^{-3}$, as in Zir2016, which is in agreement with the aforementioned determination.

\subsubsection{Number density of ISN He}
\label{sec:HeDensity}
\noindent
The unperturbed density of ISN He was determined using several methods. 

The PUI measurement by \citet{gloeckler_etal:04a} was based on the ratio between the He$^{++}$ in the core solar wind and He$^{++}$ PUIs measured by SWICS on Ulysses, independently of the instrument absolute calibration. The He density was also determined from observations of the absolute flux of He$^+$ PUIs by SWICS, but in this case the absolute calibration had to be used. The results turned out to be in excellent agreement with each other, and the density of ISN He from PUI measurements was reported at $0.0151 \pm 0.0015$~atoms~cm$^{-3}$. 

\citet{witte:04} measured the absolute density of ISN He from observations of the absolute flux of ISN He sampled at Ulysses by the GAS instrument, and obtained $0.015 \pm 0.003$~atoms~cm$^{-3}$ using the absolute calibration of the GAS instrument. Based on these measurements, \citet{mobius_etal:04a} suggested that the absolute density of ISN He in the VLISM is equal to $0.0148 \pm 0.002$~atoms~cm$^{-3}$. \citet{cummings_etal:02a} measured this density based on appropriately corrected anomalous cosmic ray fluxes and obtained $0.017 \pm 0.002$. 

In this work, we have adopted the value obtained from the PUI and direct sampling measurements: $n_{\text{He}_\text{VLISM}} = 0.0150 \pm 0.0015$~atoms~cm$^{-3}$ after \citet{gloeckler_etal:04a} and \citet{witte:04}. 

\subsubsection{Unperturbed interstellar magnetic field vector}
\label{sec:BVLISM}
\noindent
The vector of unperturbed interstellar magnetic field (ISMF) was determined based on measurements of starlight polarization on dust grains aligned with ISMF field lines and independently based on the center position and size of the IBEX Ribbon. 

The measurement based on starlight polarization obviously does not depend on any heliospheric modeling but it provides the ISMF vector averaged over several dozen parsecs. \citet{frisch_etal:15c} reported the direction of this field $(\lambda_{\text{ISMF}}, \beta_{\text{ISMF}})=(229.1\degr,41.1\degr)$ with an uncertainty of 16\degr~around this direction. The angle between this direction and the direction of Sun's motion relative to the VLISM is 43.1\degr.

\citet{grygorczuk_etal:11a, heerikhuisen_pogorelov:11a}, and Zir2016 determined the unperturbed ISMF vector by fitting the position of the IBEX Ribbon center and its diameter in the hypothesis that Ribbon is created due to the secondary ENA emission mechanism in the outer heliosheath \citep{heerikhuisen_etal:10a}. In this mechanism, the arc-like region of enhanced ENA emission marks in the sky the region where the interstellar magnetic field draped in the OHS is perpendicular to the solar-radial direction: $\vec{B} \cdot \vec{r}=0$. 

\citet{grygorczuk_etal:11a} and \citet{heerikhuisen_pogorelov:11a} adopted parameter grids with several combinations of the $\vec{B}_{\text{ISMF}}$ field strength and the inclination of $\vec{B}_{\text{ISMF}}$ to the ISN velocity $\vec{u}_{\text{VLISM}}$ and found that the best fitting parameter set is that with $\vec{B}_{\text{ISMF}} = 3\, \mu$G and the direction $(\lambda_{\text{ISMF}}, \beta_{\text{ISMF}}) = (225\degr \pm 15\degr, 35\degr \pm 5\degr)$ and $(222\degr \pm 2\degr, 41.5\degr \pm 2.5\degr)$, respectively. The resulting angles between the $\vec{B}_{\text{ISMF}}$ and $\vec{v}_{\text{VLISM}}$ vectors are 40.3\degr and 47.2\degr, respectively. 

Zir2016 used the aforementioned interstellar and solar wind parameters (except the $\vec{B}_{\text{ISMF}}$) as input to the Huntsville model of the heliosphere. They sought the strength and direction of ISMF for which the heliopause distance at the Voyager 1 direction would match that observed, the direction and strength of the draped B-field inside OHS would match that observed by Voyager 1, and the center and size of the Ribbon would be in agreement with observations of \citet{funsten_etal:13a}. They performed global simulations with different ISMF vectors, varying its strength between 2 and 4 $\mu$G, and found that the best fitting results are obtained for $B_{\text{ISMF}}=2.93 \pm 0.08$~$\mu$G, directed towards $(\lambda_{\text{ISMF}}, \beta_{\text{ISMF}}) = (227.28\degr \pm 0.69\degr, 34.62\degr \pm 0.45\degr)$. The angle between this direction and the direction of $\vec{u}_{\text{VLISM}}$ is equal to 39.5\degr, which is in agreement with the aforementioned results.

The credibility of the ISMF vector thus obtained is further enhanced by the observation by Ku2016 that the secondary population of ISN He (dubbed the Warm Breeze) flows into the heliosphere from the direction $(\lambda_{\text{WB}}, \beta_{\text{WB}}) = (251.6\degr, 12\degr)$, i.e., within the plane defined by the $\vec{u}_{\text{VLISM}}$ and $\vec{B}_{\text{ISMF}}$ vectors. State of the art global heliospheric simulations \citep[e.g., ][]{pogorelov_etal:08a,izmodenov_alexashov:15a} suggest that the ISMF direction, the Sun's motion direction, and the inflow direction of the secondary populations on the heliosphere should be co-planar. And indeed, such a geometry is implied by observations of ISN He, H, O \citep{schwadron_etal:16a}, the secondary population of ISN He, the Ribbon center  and the direction of ISMF, as illustrated in Figure~\ref{fig:NDP}.

With this, we decided to adopt the $\vec{B}_{\text{ISMF}}$ vector found by Zir2016 along with the other VLISM parameters they used in our analysis. The input VLISM parameter values are collected in Table~\ref{tab:simulParams}.

\subsection{Solar-side conditions}
\label{sec:solwindeuv}
\noindent
The solar wind flux and magnetic field contribute to the pressure balance at the heliopause and thus define the shape of the heliosphere and the distance to the heliopause. Furthermore, the charge exchange and photoionization processes together with radiation pressure modify the distribution of ISN H density inside the termination shock and hence affect the flux of pickup ions, which mediate the TS. Therefore, an appropriate model of the solar wind is essential for any global heliosphere model. Additionally, the solar EUV output is responsible for ionization of ISN He inside the TS, and thus for the ISN He signal observed by IBEX. Consequently, it is important to adopt realistic models of these factors in the global heliosphere modeling and in the model of IBEX ISN He signal. A review of the solar conditions relevant for heliospheric studies was presented by \citet{bzowski_etal:13a}. 

\subsubsection{Solar wind parameters}
\label{sec:solwind}
\noindent
In the global simulations of the heliosphere that we used, the solar wind was assumed to be time-stationary and spherically symmetric. These simplifications are justified because the most important solar wind parameter for the global shape of the heliosphere is dynamic pressure, which is close to invariant with heliolatitude \citep{mccomas_etal:08a, mccomas_etal:13b}. For the location of the heliopause, time-variations within the supersonic solar wind are of minor importance, as argued by Zir2016. In the simulations used in this paper, the boundary conditions at 1~au were adopted following Zir2016: plasma density $\rho_{\text{SW}} = 5.74 $~nuc~cm$^{-3}$, plasma temperature $T_{\text{SW}} = 51\,000$~K, solar wind speed $v_{\text{SW}} = 450$~km~s$^{-1}$, radial component of the frozen-in magnetic field $B_{r,\text{SW}} = 37.5\,\mu$G. These conditions were advected to the inner boundary of the simulation at 10~au assuming adiabatic expansion. The numerical values were chosen to obtain the heliopause distance corresponding to the heliopause crossing by Voyager 1 \citep{stone_etal:13a}. The resulting model shows an agreement with the HP crossing distance also by Voyager 2 \citep{heerikhuisen_etal:19a}. The choice of the solar wind parameters is in agreement with a reconstruction by \citet{sokol_etal:15a} based on in situ measurements \citep{king_papitashvili:05} and remote-sensing observations of interplanetary scintillations \citep{tokumaru_etal:15a}. 

\subsubsection{Ionization rates by solar EUV radiation and solar wind electron impact}
\label{sec:photoIon}
\noindent
For modeling the signal due to ISN He observed by IBEX, we adopted an observation-based model of photoionization and electron-impact losses for He inside the heliopause from \citet{bzowski_etal:13b}, extended by \citet{sokol_bzowski:14a} and \citet{sokol_etal:19a}. The dominant loss reaction is photoionization, supplemented by electron-impact ionization with the radial variation of the rate adopted from \citet{bzowski_etal:13b}. 

\section{Parameter fitting}
\label{sec:paramfitting}
\begin{figure}
\plotone{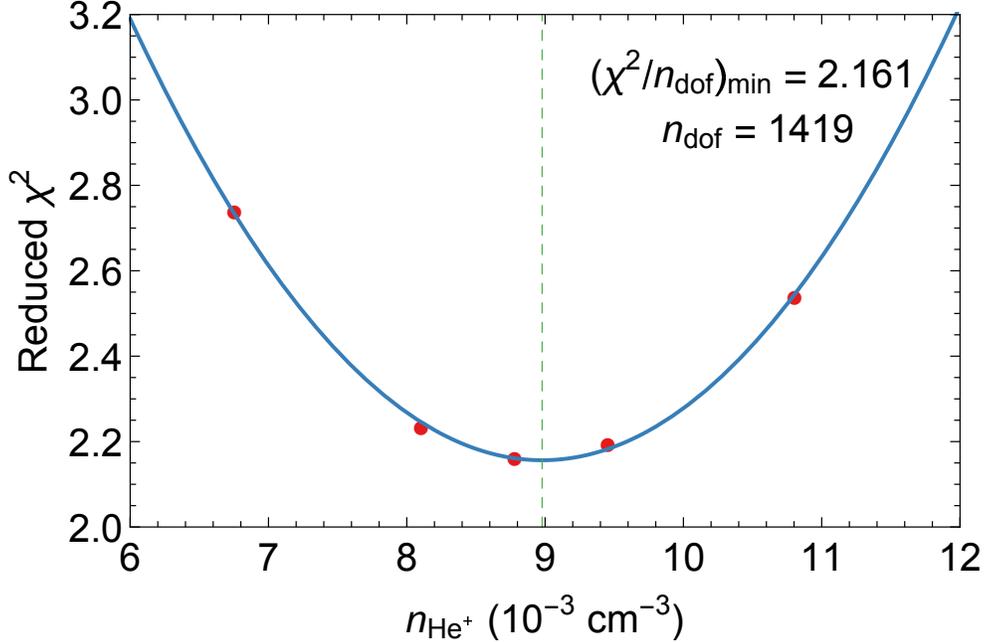}
\caption{Reduced chi-square values obtained for simulations of the signal carried out assuming various densities of He$^+$ in the VLISM. The red points show the results for the simulated values of $n_{\text{He}^+}$, and the blue line is the second order polynomial fitted to the results. The polynomial minimum is marked with the green dashed bar.}
\label{fig:chi2}
\end{figure}

\begin{sidewaysfigure}
\includegraphics[width=1.00\textheight]{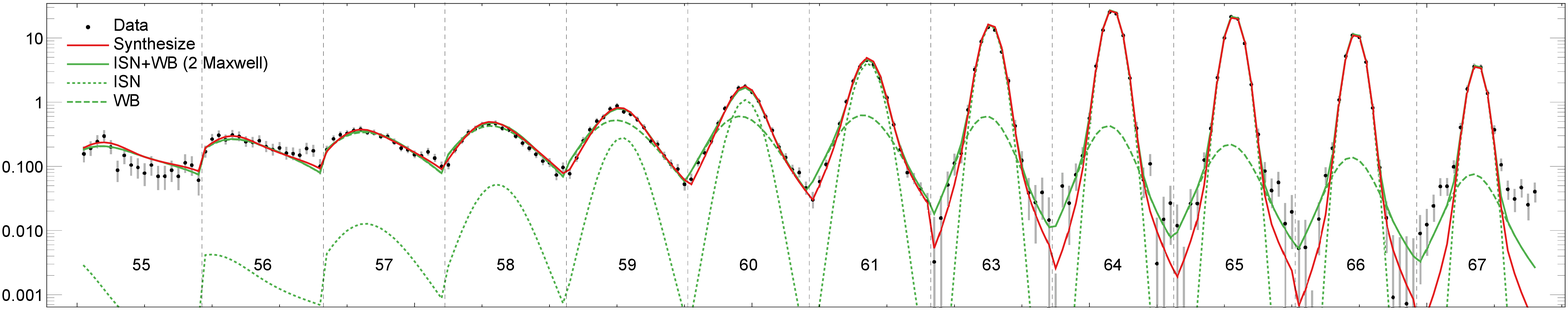}
\includegraphics[width=1.00\textheight]{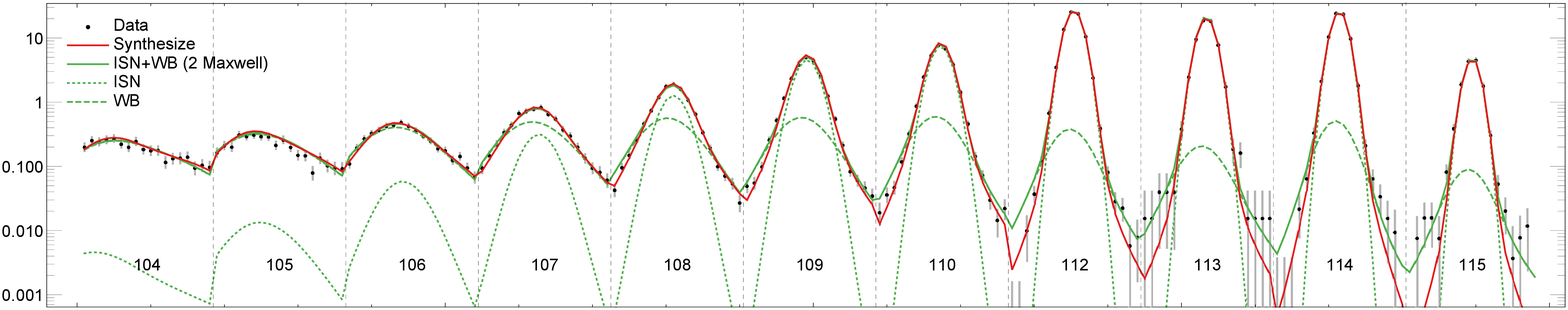}
\centering
\caption{Orbit-averaged count rates observed by IBEX-Lo and their uncertainties (black points with error bars), compared with the model obtained in this paper (red line) and the best-fitting model obtained by Ku2016 assuming that the primary and Warm Breeze (secondary) populations of ISN He are given by independent, homogeneous Maxwell-Boltzmann distributions functions in the VLISM (solid green lines). For comparison, the primary ISN He and the Warm Breeze populations from the two-Maxwellian models are shown (dotted and dashed green lines, respectively). Each panel corresponds to one observation season: 2010--2014, from top to bottom. The labels in the panels between the vertical bars indicate the reference numbers of the IBEX orbital arcs. The data between the bars are arranged by IBEX spin angle; the data cover the spin angle range from 222\degr~to 312\degr.}
\label{fig:data}
\end{sidewaysfigure}
\begin{sidewaysfigure}
\includegraphics[width=1.00\textheight]{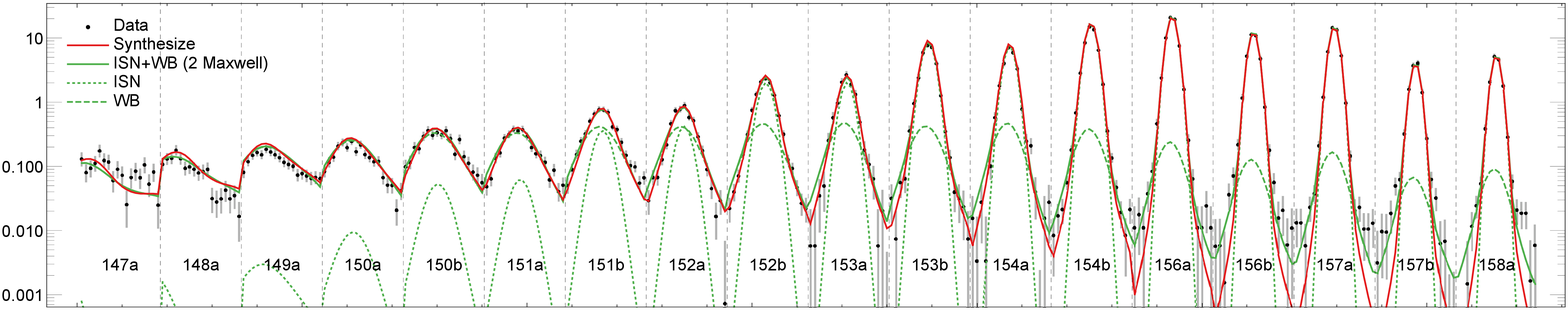}
\includegraphics[width=1.00\textheight]{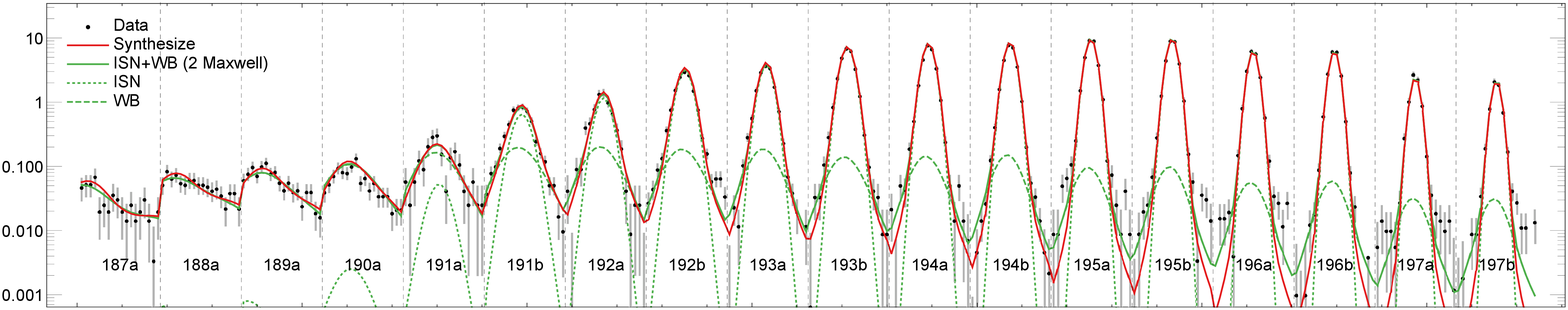}
\includegraphics[width=1.00\textheight]{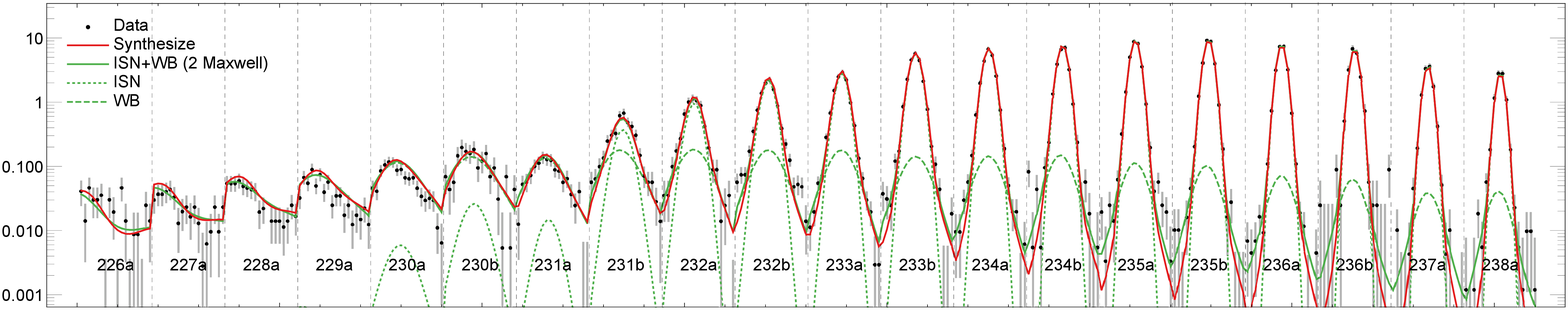}
\centering
\caption{Figure~\ref{fig:data} continued}
\label{fig:dataCont}
\end{sidewaysfigure}

\begin{sidewaysfigure}
\includegraphics[width=1.00\textheight]{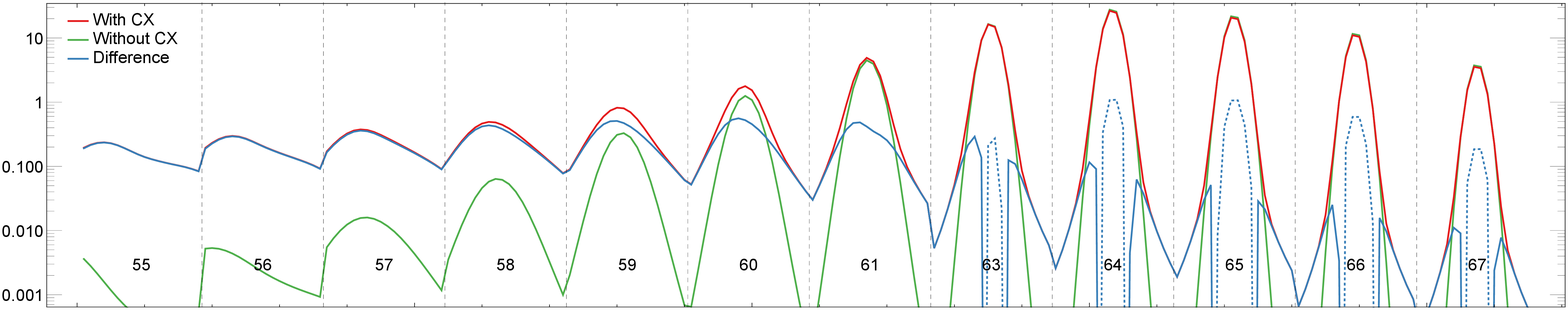}
\includegraphics[width=1.00\textheight]{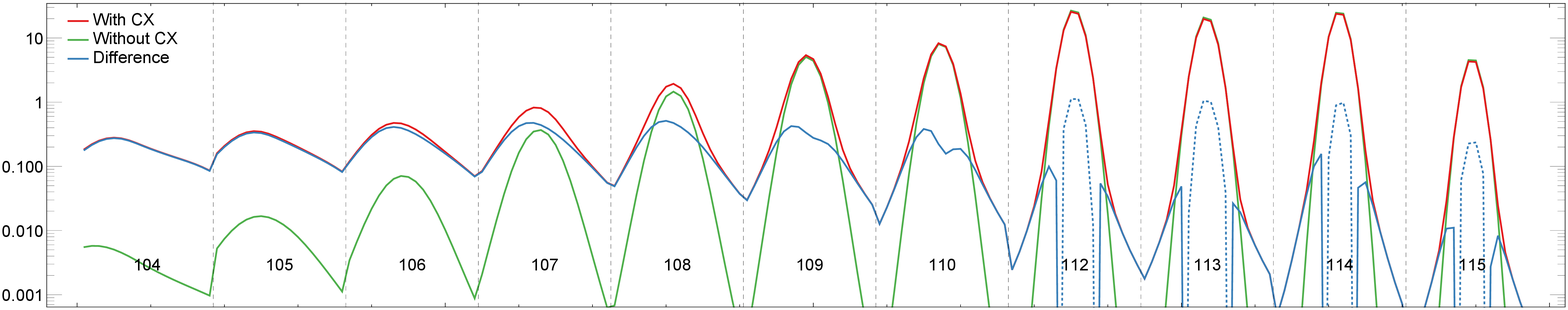}
\centering
\caption{Illustration of the filtration of the primary ISN He population in the OHS. The red line marks the result of the synthesis method, copied from Figure~\ref{fig:data}. The green line is the primary population signal obtained assuming that there are no charge-exchange processes operating in the OHS. The blue line shows the difference between these two signals. The dotted line indicates when this difference is negative. A negative difference implies that ISN atoms have been filtered out from the original unperturbed population.}
\label{fig:filtration}
\end{sidewaysfigure}
\begin{sidewaysfigure}
\includegraphics[width=1.00\textheight]{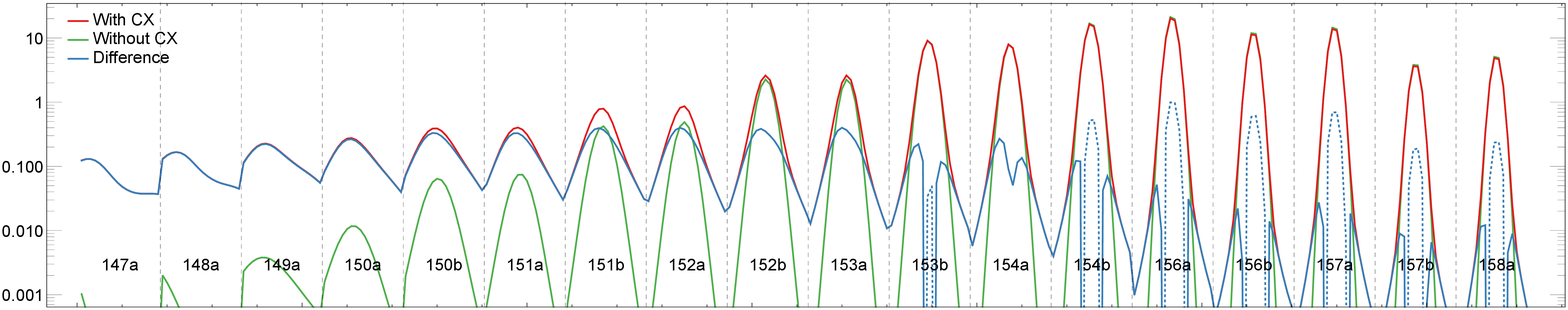}
\includegraphics[width=1.00\textheight]{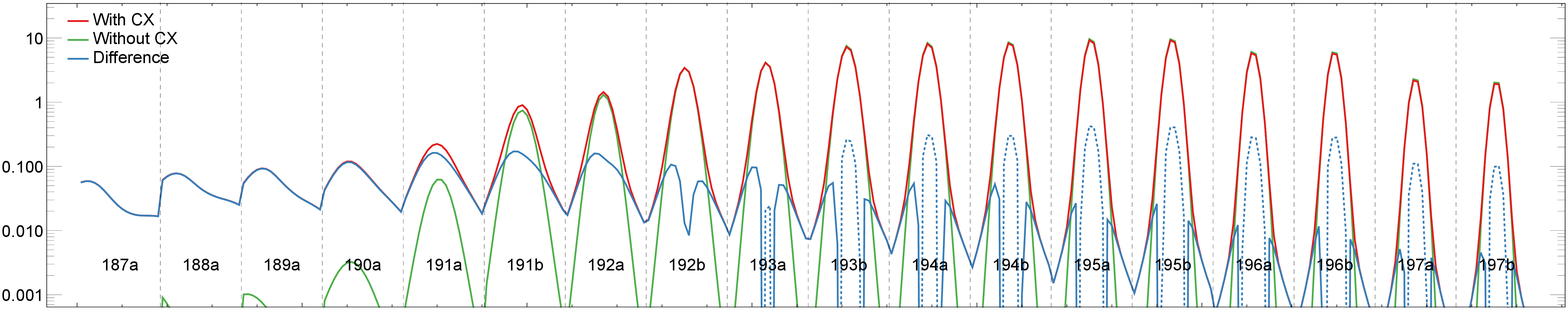}
\includegraphics[width=1.00\textheight]{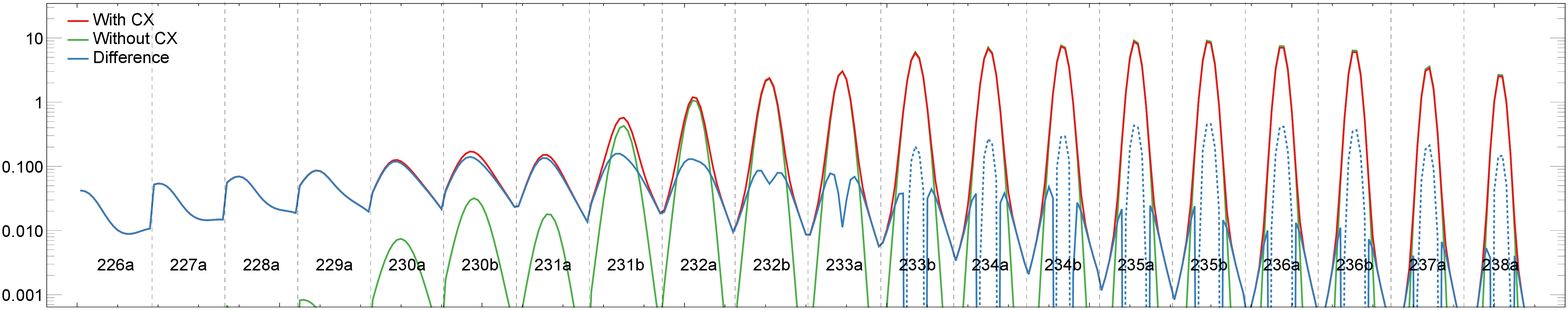}
\centering
\caption{Figure~\ref{fig:filtration} continued}
\label{fig:filtrationCont}
\end{sidewaysfigure}

\noindent
Since practically all interstellar He in the VLISM is either neutral or singly ionized \citep{slavin_frisch:08a} and other species contribute negligibly, we assumed that both in the unperturbed VLISM and in the OHS the plasma mass density $\rho_\text{pl} = (n_{\text{H}^+}+4 n_{\text{He}^+}) m_{\text{nuc}}$, where $m_{\text{nuc}}$ is nucleon mass and $n_{\text{H}^+}, n_{\text{He}^+}$ are the number densities of H$^+$ and He$^+$, respectively. With this, we calculated $n_{\text{He}^+_\text{VLISM}}$ by chi-square fitting the ISN He signal observed by IBEX from 2010 to 2014 with varying $n_{\text{He}^+_\text{VLISM}}$ while keeping $\rho_{\text{pl}_\text{VLISM}}$ unchanged. 

The observed signal was simulated for five values of $n_{\text{He}^+_\text{VLISM}}$. Subsequently, the chi-square estimator was calculated and its minimum value found. The fitted parameters included the absolute density of $n_{\text{He}^+_\text{VLISM}}$, and two instrument parameters: the conversion factor between the simulated flux and the count rate (i.e., the effective energy-independent instrument sensitivity factor), and the coefficient of reduction of instrument sensitivity for the data collected after ISN season 2012. 
The sensitivity of the IBEX-Lo instrument has been demonstrated to be very stable in time \citep{swaczyna_etal:18a}. However, due to on-board issues the post-acceleration voltage in the electrostatic analyzer section of the instrument had to be changed in 2012. This resulted in an approximately two-fold reduction of the sensitivity beginning from ISN observation season 2013. In the simulations, we assumed that the sensitivity of the instrument is a free parameter that does not depend on atom energy, and that after 2012 the instrument sensitivity is reduced by a certain factor, treated as another fitted parameter. The reduced chi-square values obtained during the fitting process are presented in Figure~\ref{fig:chi2} as a function of He$^+$ density in the VLISM. Transitioning from the simulated atom flux to count rates only involves linear scaling by certain factors. Consequently, fitting the two sensitivity parameters is done analytically for various simulated He$^+$ densities and does not require re-doing the numerical modeling of the signal. 

The uncertainty sources include (1) the statistical uncertainty due to the counting statistics, (2) background, (3) spin axis pointing, (4) deflection of the instrument optical axis from ideal alignment, (5) sensitivity to atoms with various energies, and, for the seasons 2010 to 2012, (6) the instrument throughput reduction. The only important difference is that we do not need to subtract the primary ISN He population from the signal and assess the uncertainty of this subtracted signal. 

The measurement uncertainty and the data covariance matrix were assessed identically as was done by Ku2016 based on the methodology developed by \citet{swaczyna_etal:15a}. The data used in the fitting and their uncertainties are shown in Figure~\ref{fig:data} as the red dots with error bars; note that the error bars are only approximations of the full covariance matrix of the data. The number of degrees of freedom in this study (i.e., the number of data points minus the number of fit parameters) is $N_\text{dof} =1419$. 

To find the best density of He$^+$ in the VLISM, we fit the second order polynomial to the simulated points and the minimum of the secondary polynomial is adopted as the best estimation of the density (Figure~\ref{fig:chi2}). The uncertainty of the fit is obtained from the curvature of the fit and scaled to facilitate the non-canonical value of the reduced chi-square in this study. The resulting density of He$^+$ in the unperturbed VLISM is obtained $n_{\text{He}^+_\text{VLISM}}=(8.98\pm 0.12)\times 10^{-3}$~cm$^{-3}$. The fitted energy-averaged sensitivity is $18.459\times 10^{-6}$~cm$^2$~sr, and the coefficient of the sensitivity drop due to PAC voltage reduction is 0.4633. The uncertainty of $n_{\text{He}^+_\text{,VLISM}}$ density quoted above is solely the fit uncertainty. The best-fit model is presented along with the data with red line in Figure~\ref{fig:data}. 

The minimum chi-square per degree of freedom (i.e., the minimum value of reduced chi-square) obtained from the minimization is equal to 2.161, while the expected value is $1 \pm (2/N_{\text{dof}})^{1/2} = 1 \pm 0.038$. This implies that the model is not perfect. Nevertheless, the agreement between the data and our present model is better than that obtained for the model with two independent Maxwell-Boltzmann populations as in Ku2016. The chi-square value calculated for the data set that has been updated to the identical set used with our new model by evaluating the two-Maxwellian model with the parameters of the primary ISN He from \citet{bzowski_etal:15a} and the secondary population parameters reported by Ku2016 is 2.204 (green lines in Figure~\ref{fig:data}). Importantly, we could only minimize one parameter of the physical system, i.e., the density of He$^+$, and the two parameters of the instrument sensitivity, leaving all the other parameters of the problem fixed. By contrast, in the approach adopted by Ku2016, even though the physical model was simpler than ours, the number of free parameters in the fit was larger. With a larger number of fit parameters the chi-square values tend to decrease. In our case, we have obtained a lower chi-square value, which suggests that the model presented in this paper is closer to the physical reality than the two-Maxwellian model used by Ku2016. Nonetheless, the systematic discrepancies between the measured and modeled count rates in the wings of later orbits during each season are generally larger for the current model compared to the two-Maxwellian model. This discrepancy may suggest that the unperturbed ISN He population is not fully equilibrated and thus it may be better described by the kappa distribution far from the heliosphere instead of the Maxwell distribution \citep{sokol_etal:15a, swaczyna_etal:19a}. 

Figure~\ref{fig:filtration} presents a comparison between the signal from our full synthesis method and the uperturbed ISN He population with the source region set to 1000~au. As evident from this comparison, the primary ISN He population observed by IBEX is not pristine because it has been modified by filtration. We surmise that the ``filtered'' portion of the signal, i.e., the negative difference between the synthesis method and the one-Maxwellian model with the filtration effects neglected, may masquerade in the comparison of data with the two-Maxwellian model as residuals that resemble patterns characteristic for a kappa distribution function of the unperturbed ISN He. More in-depth investigation of this aspect will be the subject of future studies.

To further assess the superiority of the present model over the two-Maxwellian approximation we calculated chi-square on a subset of data where the Warm Breeze dominates. We used data from orbits 055--061, 104--109, 150a--153a, 187a--192b, and 226a-233a, all together 756 data points. For the synthesis method, we obtained chi-squared equal to 1516.24, and for the two-Maxwellian model to 1529.75. Reduced chi-squared values were equal to 2.013 and 2.032, respectively. This suggests that also for the orbits where little of the primary population is expected, the present model fits a little better than the two-Maxwellian approximation. However, when one includes all data used in our fitting except the bins used to fit the inflow parameters of the primary population by \citet{bzowski_etal:15a}, one obtains a reduced chi-square equal to 2.158 for the two-Maxwellian model and to 2.167 for the synthesis method. We speculate that this is because of a contribution to the data from ISN H, which was not subtracted from the data. Alteratively, it may be due to the fact that the ISN He inflow parameters we use here slightly differ from the optimum fit obtained by \citet{bzowski_etal:15a}.  

The method of fitting He$^+$ density in the VLISM that we have used gives best results when the consistency of the model used to calculate the statistical weights $\omega$ with the global heliosphere simulation model is maintained. The quality of the fitting is very sensitive to this aspect. When in the synthesis method we assume the parameters of ISN He from \citet{bzowski_etal:15a} instead of those from \citet{mccomas_etal:15b}, but keep unchanged the parameters in the global heliosphere model (Zir2016 used the parameters from \citet{mccomas_etal:15b}), then we introduce a slight inconsistency into the simulation. We assume that ISN He is flowing a little differently than the plasma at the boundary of the simulation system. From that, we obtained in the fitting the same magnitude of He$^+$ density (within the fit uncertainty), but a statistically significantly larger chi-squared value of 2.43, larger than in our best fit. In fact, it is larger than the chi-square value calculated for the two-Maxwellian model with the parameters from \citet{bzowski_etal:15a} and Ku2016. With this, the interpretation of the present model as superior to the two-Maxwellian model would not have been justified.  

We consider this requirement of high level of self-consistency in the modeling as a major strength of our approach and the result. While the inferred density of He$^+$ is model dependent, the model we used is self-consistent. Consequently, it is not advisable to modify one of its aspects (e.g., heliopause location; B field direction or strength; allowing for kappa distribution functions for the plasma or ISN He; tensor-like thermal spread of the plasma or ISN gas, etc.) without propagating it self-consistently into the global heliosphere model and the statistical weight calculation. This makes it a suitable tool for verifying predictions of various global heliosphere models against observations, with different approximations made or different parameters used. 

\section{VLISM parameter derivation}
\label{sec:VLISMParamDeriv}

\begin{deluxetable}{rc}
\tablecaption{VLISM parameter values}
\tablehead{
\colhead{quantity}  &
\colhead{magnitude}
}
\startdata
He$^+$ number density $n_{\text{He}^+}$  & $(8.98 \pm 0.12)\times 10^{-3}$~cm$^{-3}$    \\
proton number density $n_{\text{H}^+}$   & $5.41\times 10^{-2}$~cm$^{-3}$    \\
electron number density $n_\text{e}$     & $6.30 \times 10^{-2}$~cm$^{-3}$   \\
hydrogen ionization degree $X_\text{H}$  & $0.26$\tablenotemark{a}            \\
helium ionization degree $X_\text{He}$   & $0.37$                            \\
\enddata
\tablenotetext{a}{For $n_H = 0.154$~cm$^{-3}$, adopted by Zir2016. If $n_\text{H} = 0.16$~cm$^{-3}$ is adopted as measured by \citet{bzowski_etal:09a}, then $X_\text{H} = 0.25$ is received, which is very close to the previous value. }
\label{tab:VLISMParamsResults}
\end{deluxetable}

\noindent 
Using the density of interstellar He$^+$ component obtained from fitting and with the model parameters listed in Table~\ref{tab:simulParams}, we derive other parameters of the VLISM as follows. To calculate the proton density, we do not need to assume the frequently used cosmological He/H abundance because having all relevant quantities measured we do not need to pre-assume anything in this respect. Generally, the cosmological abundance is not a reliable estimate for the local He/H abundance because of different chemical processing of matter in various populations of stars and its subsequent redistribution by supernova explosions in different regions of space \citep[e.g.,][]{wilson_rood:94a}. In particular, the Local Bubble and the Local Interstellar Medium very likely were heavily processed by a series of supernova explosions a few million years ago \citep{breitschwerdt_etal:96a}. With the mass density of the VLISM plasma 
\begin{equation}
\rho_{\text{pl}} =(n_{\text{H}^+} + 4\, n_{\text{He}^+}) m_{\text{nuc}}
\label{eq:plasmaDensDef}
\end{equation}
we calculate
\begin{equation}
n_{\text{H}^+}  =  \rho_\text{pl}/m_{\text{nuc}} - 4 n_{\text{He}^+}.
\label{eq:nH+Dens} 
\end{equation}

The electron density in the VLISM is given by:
\begin{equation}
n_\text{e} = n_{\text{H}^+} + n_{\text{He}^+},
\label{eq:nElDens}
\end{equation}
the ionization degree of He in the VLISM is equal to
\begin{equation}
X_{\text{He}} = n_{\text{He}^+}/(n_{\text{He}^+} + n_{\text{He}}), 
\label{eq:HeIonDeg}
\end{equation}
and the ionization degree of ISN H is obtained as 
\begin{equation}
X_{\text{H}} = n_{\text{H}^+}/(n_\text{H} + n_{\text{H}^+}). 
\label{eq:ionDegHe}
\end{equation}
Numerical values for these parameters are listed in Table~\ref{tab:VLISMParamsResults}. The densities of the ionized components we have obtained do not imply a deviation from the cosmological H/He ratio larger than $\sim 15$\%.

\section{Discussion and conclusions}
\label{sec:discuss}
\noindent
The data we have used were collected between November 2009 and March 2014, over a time span of $\sim 4.5$~years (with 8-month gaps in measurements each year). The speed of the Sun's motion relative to the VLISM is 25.4~km~s$^{-1}$, i.e., 5.36~au~year$^{-1}$; thus, during the measurement interval the Sun has covered less than $23\,\text{au} = 1.1\times 10^{-4}$ parsecs relative to this matter, which is at least 5 orders of magnitude less than the expected size of the cloud of interstellar matter the Sun is traversing. In the magnetic field determination performed by Zir2016 based on fitting the Ribbon size and location in the sky, the ENA signal originates within $\sim 500$~au from the Sun (cf Figure 1 in their paper), i.e., $2.5 \times 10^{-3}$~pc. Therefore, from the perspective of the size of a parsec, typical for interstellar clouds in the Sun's neighborhood, the result of this analysis can be regarded as a point measurement of the plasma condition in the VLISM.  

\citet{slavin_frisch:08a} performed a parametric study of the VLISM conditions based on radiative transfer calculations and the data available back then, in particular for the previously thought VLISM temperature of 6300~K. The goal was to match the available data on the Local Interstellar Cloud (LIC) including column densities of several ions based on absorption lines and \emph{in situ} data such as the neutral He density. This was done by constructing the ionizing radiation field based on (1) directly observed nearby hot stars and (2) modeled emission from hot gas in the Local Bubble \citep[responsible for the diffuse soft X-ray background, ][]{mccammon_etal:83a,snowden_etal:97a}. An additional component to the radiation field from a hypothesized evaporative boundary to the LIC was also modeled and included. The radiation field was transferred through the cloud to the location of the Sun using the radiative transfer/thermal equilibrium code Cloudy \citep{ferland_etal:13a}. Various parameters, such as elemental abundances and the density, and magnetic field in the cloud, were then varied to achieve good matches to the data.

\citet{slavin_frisch:08a} explored a large grid of parameters (42, see their Table 2 for the input values and Table 4 for the results), and they isolated several of the most promising sets. Since then, estimates for the VLISM temperature have increased to $\sim 7500$ K and the strength of magnetic field has been better constrained $(\sim 3\,\mu$G). It has also become clear that a fraction of the emission in the soft X-ray diffuse background originates within the heliosphere, generated by charge exchange between inflowing neutrals and solar wind ions \citep{snowden:15a}. As a result, the intensity of the EUV/X-ray emission from the hot gas of the Local Bubble was overestimated in \citet{slavin_frisch:08a}. Recent estimates of the fraction of the soft X-ray emission coming from the Local Bubble near the Galactic plane range from 26\% \citep{smith_etal:14a} to 60\% \citep{galeazzi_etal:14a}. The Cloudy code has also been improved in the intervening years. 

Given these changes, we have re-calculated the ionization of the LIC for several parameter sets. The major assumed parameters in this modeling are the temperature of the hot gas in the Local Bubble and the opacity of the cloud, which depends on the column density. The cloud density and the cloud magnetic field were varied to achieve the desired values for the neutral He density and temperature at the heliosphere. In addition, we have looked at different fractions of the soft X-rays that come from the Local Bubble. Since the soft X-ray background is brightest out of the Galactic plane, the fractions mentioned above are lower limits. For an assumed fraction of 75\%, a Local Bubble temperature of $10^6$~K, and cloud column density of $N($\ion{H}{1}$) = 4\times10^{17}$~cm$^{-2}$, we found we need a magnetic field of 3.5 $\mu$G. The values found for ionization then are $X_{\text{H}^+} = 0.245$, $X_{\text{He}^+} = 0.395$, $n_{\text{H}^+} = 6.2\times 10^{-2}$~cm$^{-3}$, $n_{\text{He}^+} = 9.9 \times10^{-3}$~cm$^{-3}$, and $n_e = 7.2\times 10^{-2}$~cm$^{-3}$. Using 50\% for the soft X-ray fraction yields similar results but requires a higher magnetic field, $B = 4\,\mu$G. These results had been generated before the presently reported IBEX-Lo results were available. 

\citet{wolff_etal:99}, based on EUVE observations of nearby white dwarfs, compared column densities of H, He, and He$^+$ in the Local Interstellar Medium and found that the ionization degree of He is 0.4 (with a large uncertainty), consistent with our value of 0.37, and the He$^+$/H ratio they obtained is $0.052 \pm 0.007$, in agreement with our 0.056 -- 0.058. This suggests that the properties of VLISM obtained independently from astrophysical and heliospheric observations converge.

Making discrimination between alternative VLISM parameter sets solely with the use of radiative transfer calculations and telescopic observations seems challenging. Our analysis suggests that observations of ISN He can help obtain this discrimination. Our fitting the ISN He observations from IBEX-Lo resulted in an assessment for the density of He$^+$ $8.98\times 10^{-3} $~cm$^{-3}$, proton density $5.41 \times 10^{-2}$~cm$^{-3}$, electron density $6.30 \times 10^{-2}$~cm$^{-3}$, ionization degrees of H and He $0.26$ and  $0.37$, respectively. This is in very good agreement with the aforementioned estimates based on equilibrium models of the VLISM. This agreement lends credence to the consistency of the global heliosphere model and the physical state of the VLISM.

\acknowledgments
M.B., A.C., and E.J.Z. gratefully acknowledge the collaboration within ISSI Team 368 \emph{The Physics of the Very Local Interstellar Medium and its interaction with the heliosphere}. The work at CBK PAN was supported by the Polish National Science Centre grants 2015/18/M/ST9/00036 and 2015/19/B/ST9/01328. J.H. and E.J.Z. acknowledge support from NASA grant 80NSSC18K1212. The US portion of this work was funded by the IBEX mission as a part of the NASA Explorer Program (NNG17FC93C; NNX17AB04G). E.J.Z. also acknowledges partial support from NASA grant 80NSSC17K0597. A.G. and P.W. were supported by the Swiss national Science Foundation.

\bibliographystyle{aasjournal}
\bibliography{iplbib}

\begin{thebibliography}{}
\expandafter\ifx\csname natexlab\endcsname\relax\def\natexlab#1{#1}\fi
\providecommand{\url}[1]{\href{#1}{#1}}
\providecommand{\dodoi}[1]{doi:~\href{http://doi.org/#1}{\nolinkurl{#1}}}
\providecommand{\doeprint}[1]{\href{http://ascl.net/#1}{\nolinkurl{http://ascl.net/#1}}}
\providecommand{\doarXiv}[1]{\href{https://arxiv.org/abs/#1}{\nolinkurl{https://arxiv.org/abs/#1}}}

\bibitem[{{Adams} \& {Frisch}(1977)}]{adams_frisch:77a}
{Adams}, T.~F., \& {Frisch}, P.~C. 1977, \apj, 212, 300, \dodoi{10.1086/155048}

\bibitem[{Baranov \& Malama(1993)}]{baranov_malama:93}
Baranov, V.~B., \& Malama, Y.~G. 1993, \jgr, 98, 15157

\bibitem[{Barnett {et~al.}(1990)Barnett, Hunter, Kirkpatrick, Alvarez,
  Cisneros, \& Phaneuf}]{barnett_etal:90}
Barnett, C.~F., Hunter, H.~T., Kirkpatrick, M.~I., {et~al.} 1990, Atomic data
  for fusion. Collisions of {H}, {H}$_2$, {H}e and {L}i atoms and ions with
  atoms and molecules, Vol. ORNL-6086/V1 (Oak Ridge, Tenn.: Oak Ridge National
  Laboratories)

\bibitem[{{Bloch} {et~al.}(1986){Bloch}, {Jahoda}, {Juda}, {McCammon},
  {Sanders}, \& {Snowden}}]{bloch_etal:86a}
{Bloch}, J.~J., {Jahoda}, K., {Juda}, M., {et~al.} 1986, \apjl, 308, L59,
  \dodoi{10.1086/184744}

\bibitem[{{Breitschwerdt} {et~al.}(1996){Breitschwerdt}, {Egger}, {Freyberg},
  {Frisch}, \& {Vallerga}}]{breitschwerdt_etal:96a}
{Breitschwerdt}, D., {Egger}, R., {Freyberg}, M.~J., {Frisch}, P.~C., \&
  {Vallerga}, J.~V. 1996, \ssr, 78, 183, \dodoi{10.1007/BF00170805}

\bibitem[{{Burlaga} \& {Ness}(2016)}]{burlaga_ness:16a}
{Burlaga}, L.~F., \& {Ness}, N.~F. 2016, \apj, 829, 134,
  \dodoi{10.3847/0004-637X/829/2/134}

\bibitem[{{Bzowski} {et~al.}(2017){Bzowski}, {Kubiak}, {Czechowski}, \&
  {Grygorczuk}}]{bzowski_etal:17a}
{Bzowski}, M., {Kubiak}, M.~A., {Czechowski}, A., \& {Grygorczuk}, J. 2017,
  \apj, 845, 15.
\newblock \doarXiv{1707.02193}

\bibitem[{{Bzowski} {et~al.}(2014){Bzowski}, {Kubiak}, {H{\l}ond},
  {Sok{\'o}{\l}}, {Banaszkiewicz}, \& {Witte}}]{bzowski_etal:14a}
{Bzowski}, M., {Kubiak}, M.~A., {H{\l}ond}, M., {et~al.} 2014, \aap, 569, A8,
  \dodoi{10.1051/0004-6361/201424127}

\bibitem[{{Bzowski} {et~al.}(2008){Bzowski}, {M{\"o}bius}, {Tarnopolski},
  {Izmodenov}, \& {Gloeckler}}]{bzowski_etal:08a}
{Bzowski}, M., {M{\"o}bius}, E., {Tarnopolski}, S., {Izmodenov}, V., \&
  {Gloeckler}, G. 2008, \aap, 491, 7, \dodoi{10.1051/0004-6361:20078810}

\bibitem[{{Bzowski} {et~al.}(2009){Bzowski}, {M{\"o}bius}, {Tarnopolski},
  {Izmodenov}, \& {Gloeckler}}]{bzowski_etal:09a}
---. 2009, \ssr, 143, 177, \dodoi{10.1007/s11214-008-9479-0}

\bibitem[{Bzowski {et~al.}(2013{\natexlab{a}})Bzowski, Sok{\'o}{\l}, Kubiak, \&
  Kucharek}]{bzowski_etal:13b}
Bzowski, M., Sok{\'o}{\l}, J.~M., Kubiak, M.~A., \& Kucharek, H.
  2013{\natexlab{a}}, \aap, 557, A50, \dodoi{10.1051/0004-6361/201321700}

\bibitem[{Bzowski {et~al.}(2012)Bzowski, Kubiak, M{\"o}bius, Bochsler, Leonard,
  Heirtzler, Kucharek, Sok{\'{o}}{\l}, H{\l}ond, Crew, Schwadron, Fuselier, \&
  McComas}]{bzowski_etal:12a}
Bzowski, M., Kubiak, M.~A., M{\"o}bius, E., {et~al.} 2012, \apjs, 198, 12,
  \dodoi{10.1088/0067-0049/198/2/12}

\bibitem[{Bzowski {et~al.}(2013{\natexlab{b}})Bzowski, Sok{\'{o}}{\l},
  Tokumaru, Fujiki, Qu{\'e}merais, Lallement, Ferron, Bochsler, \&
  McComas}]{bzowski_etal:13a}
Bzowski, M., Sok{\'{o}}{\l}, J.~M., Tokumaru, M., {et~al.} 2013{\natexlab{b}},
  in {Cross-Calibration of Far {UV} Spectra of Solar Objects and the
  Heliosphere}, ed. E.~Qu{\'e}merais, M.~Snow, \& R.~Bonnet, {ISSI Scientific
  Report} No.~13 ({Springer Science+Business Media}), 67--138

\bibitem[{{Bzowski} {et~al.}(2015){Bzowski}, {Swaczyna}, {Kubiak},
  {Sok\'{o}{\l}}, {Fuselier}, {Galli}, {Heirtzler}, {Kucharek}, {Leonard},
  {McComas}, {M{\"o}bius}, {Schwadron}, \& {Wurz}}]{bzowski_etal:15a}
{Bzowski}, M., {Swaczyna}, P., {Kubiak}, M., {et~al.} 2015, \apjs, 220, 28,
  \dodoi{10.1088/0067-0049/220/2/28}

\bibitem[{{Crutcher}(1982)}]{crutcher:82a}
{Crutcher}, R.~M. 1982, \apj, 254, 82, \dodoi{10.1086/159707}

\bibitem[{{Cummings} {et~al.}(2002){Cummings}, {Stone}, \&
  {Steenberg}}]{cummings_etal:02a}
{Cummings}, A.~C., {Stone}, E.~C., \& {Steenberg}, C.~D. 2002, \apj, 578, 194,
  \dodoi{10.1086/342427}

\bibitem[{Fahr \& Mueller(1967)}]{fahr_mueller:67}
Fahr, H.~J., \& Mueller, K.~G. 1967, Z. Phys., 200, 343

\bibitem[{Fahr \& Ruci{\'n}ski(1999)}]{fahr_rucinski:99}
Fahr, H.~J., \& Ruci{\'n}ski, D. 1999, \aap, 350, 1071

\bibitem[{{Ferland} {et~al.}(2013){Ferland}, {Porter}, {van Hoof}, {Williams},
  {Abel}, {Lykins}, {Shaw}, {Henney}, \& {Stancil}}]{ferland_etal:13a}
{Ferland}, G.~J., {Porter}, R.~L., {van Hoof}, P.~A.~M., {et~al.} 2013, \rmxaa,
  49, 137.
\newblock \doarXiv{1302.4485}

\bibitem[{{Frisch} {et~al.}(2002){Frisch}, {Grodnicki}, \&
  {Welty}}]{frisch_etal:02a}
{Frisch}, P.~C., {Grodnicki}, L., \& {Welty}, D.~E. 2002, \apj, 574, 834,
  \dodoi{10.1086/341001}

\bibitem[{{Frisch} {et~al.}(2011){Frisch}, {Redfield}, \&
  {Slavin}}]{frisch_etal:11a}
{Frisch}, P.~C., {Redfield}, S., \& {Slavin}, J.~D. 2011, \araa, 49, 237,
  \dodoi{10.1146/annurev-astro-081710-102613}

\bibitem[{{Frisch} {et~al.}(2015){Frisch}, {Berdyugin}, {Piirola}, {Magalhaes},
  {Seriacopi}, {Wiktorowicz}, {Andersson}, {Funsten}, {McComas}, {Schwadron},
  {Slavin}, {Hanson}, \& {Fu}}]{frisch_etal:15c}
{Frisch}, P.~C., {Berdyugin}, A., {Piirola}, V., {et~al.} 2015, \apj, 814, 112,
  \dodoi{10.1088/0004-637X/814/2/112}

\bibitem[{{Funsten} {et~al.}(2013){Funsten}, {DeMajistre}, {Frisch},
  {Heerikhuisen}, {Higdon}, {Janzen}, {Larsen}, {Livadiotis}, {McComas},
  {M{\"o}bius}, {Reese}, {Reisenfeld}, {Schwadron}, \&
  {Zirnstein}}]{funsten_etal:13a}
{Funsten}, H.~O., {DeMajistre}, R., {Frisch}, P.~C., {et~al.} 2013, \apj, 776,
  30, \dodoi{10.1088/0004-637X/776/1/30}

\bibitem[{{Fuselier} {et~al.}(2009){Fuselier}, {Bochsler}, {Chornay}, {Clark},
  {Crew}, {Dunn}, {Ellis}, {Friedmann}, {Funsten}, {Ghielmetti}, {Googins},
  {Granoff}, {Hamilton}, {Hanley}, {Heirtzler}, {Hertzberg}, {Isaac}, {King},
  {Knauss}, {Kucharek}, {Kudirka}, {Livi}, {Lobell}, {Longworth}, {Mashburn},
  {McComas}, {M{\"o}bius}, {Moore}, {Moore}, {Nemanich}, {Nolin}, {O'Neal},
  {Piazza}, {Peterson}, {Pope}, {Rosmarynowski}, {Saul}, {Scherrer}, {Scheer},
  {Schlemm}, {Schwadron}, {Tillier}, {Turco}, {Tyler}, {Vosbury}, {Wieser},
  {Wurz}, \& {Zaffke}}]{fuselier_etal:09b}
{Fuselier}, S.~A., {Bochsler}, P., {Chornay}, D., {et~al.} 2009, \ssr, 146,
  117, \dodoi{10.1007/s11214-009-9495-8}

\bibitem[{Galeazzi {et~al.}(2014)Galeazzi, Chiao, Collier, Cravens, Koutroumpa,
  Kuntz, Lallement, Lepri, McCammon, Morgan, Porter, Robertson, Snowden,
  Thomas, Uprety, Ursino, \& Walsh}]{galeazzi_etal:14a}
Galeazzi, M., Chiao, M., Collier, M.~R., {et~al.} 2014, \nat, 512, 171,
  \dodoi{10.1038/nature13525}

\bibitem[{{Galli} {et~al.}(2015){Galli}, {Wurz}, {Park}, {Kucharek},
  {M{\"o}bius}, {Schwadron}, {Sok\'{o}{\l}}, {Bzowski}, {Kubiak}, {Swaczyna},
  {Fuselier}, \& {McComas}}]{galli_etal:15a}
{Galli}, A., {Wurz}, P., {Park}, J., {et~al.} 2015, \apjs, 220, 30,
  \dodoi{10.1088/0067-0049/220/2/30}

\bibitem[{Galli {et~al.}(2016)Galli, Wurz, Schwadron, Kucharek, M{\"o}bius,
  Bzowski, Sok{\'o}{\l}, Kubiak, Funsten, Fuselier, \&
  McComas}]{galli_etal:16a}
Galli, A., Wurz, P., Schwadron, N., {et~al.} 2016, \apj, 821, 107,
  \dodoi{10.3847/0004-637X/821/2/107}

\bibitem[{{Galli} {et~al.}(2019){Galli}, {Wurz}, {Rahmanifard}, {M{\"o}bius},
  {Schwadron}, {Kucharek}, {Heirtzler}, {Fairchild}, {Bzowski}, {Kubiak},
  {Kowalska-Leszczy{\'n}ska}, {Sok{\'o}{\l}}, {Fuselier}, {Swaczyna}, \&
  {McComas}}]{galli_etal:19a}
{Galli}, A., {Wurz}, P., {Rahmanifard}, F., {et~al.} 2019, \apj, 871, 52,
  \dodoi{10.3847/1538-4357/aaf737}

\bibitem[{{Gloeckler} {et~al.}(2004){Gloeckler}, {Allegrini}, {Elliott},
  {McComas}, {Schwadron}, {Geiss}, {von Steiger}, \&
  {Jones}}]{gloeckler_etal:04a}
{Gloeckler}, G., {Allegrini}, F., {Elliott}, H.~A., {et~al.} 2004, \apjl, 604,
  L121

\bibitem[{{Gloeckler} \& {Geiss}(2001)}]{gloeckler_geiss:01a}
{Gloeckler}, G., \& {Geiss}, J. 2001, \ssr, 97, 169

\bibitem[{{Gry} \& {Jenkins}(2014)}]{gry_jenkins:14a}
{Gry}, C., \& {Jenkins}, E.~B. 2014, \aap, 567, A58,
  \dodoi{10.1051/0004-6361/201323342}

\bibitem[{{Grygorczuk} {et~al.}(2011){Grygorczuk}, {Ratkiewicz}, {Strumik}, \&
  {Grzedzielski}}]{grygorczuk_etal:11a}
{Grygorczuk}, J., {Ratkiewicz}, R., {Strumik}, M., \& {Grzedzielski}, S. 2011,
  \apjl, 727, L48, \dodoi{10.1088/2041-8205/727/2/L48}

\bibitem[{Gurnett {et~al.}(2013)Gurnett, Kurth, Burlaga, \&
  Ness}]{gurnett_etal:13a}
Gurnett, D.~A., Kurth, W.~S., Burlaga, L.~F., \& Ness, N.~F. 2013, Science,
  341, 1489, \dodoi{10.1126/science.1241681}

\bibitem[{{Gurnett} {et~al.}(2015){Gurnett}, {Kurth}, {Stone}, {Cummings},
  {Krimigis}, {Decker}, {Ness}, \& {Burlaga}}]{gurnett_etal:15a}
{Gurnett}, D.~A., {Kurth}, W.~S., {Stone}, E.~C., {et~al.} 2015, \apj, 809,
  121, \dodoi{10.1088/0004-637X/809/2/121}

\bibitem[{{Heerikhuisen} \& {Pogorelov}(2010)}]{heerikhuisen_pogorelov:10a}
{Heerikhuisen}, J., \& {Pogorelov}, N.~V. 2010, in Astronomical Society of the
  Pacific Conference Series, Vol. 429, Numerical Modeling of Space Plasma
  Flows, Astronum-2009, ed. N.~V. {Pogorelov}, E.~{Audit}, \& G.~P. {Zank},
  227--232

\bibitem[{{Heerikhuisen} \& {Pogorelov}(2011)}]{heerikhuisen_pogorelov:11a}
{Heerikhuisen}, J., \& {Pogorelov}, N.~V. 2011, \apj, 738, 29,
  \dodoi{10.1088/0004-637X/738/1/29}

\bibitem[{{Heerikhuisen} {et~al.}(2015){Heerikhuisen}, {Zirnstein}, \&
  {Pogorelov}}]{heerikhuisen_etal:15a}
{Heerikhuisen}, J., {Zirnstein}, E., \& {Pogorelov}, N. 2015, \jgr, 120, 1516,
  \dodoi{10.1002/2014JA020636}

\bibitem[{{Heerikhuisen} {et~al.}(2019){Heerikhuisen}, {Zirnstein},
  {Pogorelov}, {Zank}, \& {Desai}}]{heerikhuisen_etal:19a}
{Heerikhuisen}, J., {Zirnstein}, E.~J., {Pogorelov}, N.~V., {Zank}, G.~P., \&
  {Desai}, M. 2019, \apj, 874, 76, \dodoi{10.3847/1538-4357/ab05e3}

\bibitem[{{Heerikhuisen} {et~al.}(2010){Heerikhuisen}, {Pogorelov}, {Zank},
  {Crew}, {Frisch}, {Funsten}, {Janzen}, {McComas}, {Reisenfeld}, \&
  {Schwadron}}]{heerikhuisen_etal:10a}
{Heerikhuisen}, J., {Pogorelov}, N.~V., {Zank}, G.~P., {et~al.} 2010, \apjl,
  708, L126, \dodoi{10.1088/2041-8205/708/2/L126}

\bibitem[{Isenberg(1986)}]{isenberg:86}
Isenberg, P.~A. 1986, \jgr, 91, 9965

\bibitem[{{Izmodenov} {et~al.}(2003{\natexlab{a}}){Izmodenov}, {Gloeckler}, \&
  {Malama}}]{izmodenov_etal:03b}
{Izmodenov}, V., {Gloeckler}, G., \& {Malama}, Y. 2003{\natexlab{a}}, \grl, 30,
  3

\bibitem[{{Izmodenov} {et~al.}(2003{\natexlab{b}}){Izmodenov}, {Malama},
  {Gloeckler}, \& {Geiss}}]{izmodenov_etal:03a}
{Izmodenov}, V., {Malama}, Y.~G., {Gloeckler}, G., \& {Geiss}, J.
  2003{\natexlab{b}}, \apjl, 594, L59

\bibitem[{Izmodenov \& Alexashov(2015)}]{izmodenov_alexashov:15a}
Izmodenov, V.~V., \& Alexashov, D.~B. 2015, \apjs, 220, 32,
  \dodoi{10.1088/0067-0049-220-2-32}

\bibitem[{{King} \& {Papitashvili}(2005)}]{king_papitashvili:05}
{King}, J.~H., \& {Papitashvili}, N.~E. 2005, \jgr, 110, 2104,
  \dodoi{10.1029/2004JA010649}

\bibitem[{{Kowalska-Leszczynska} {et~al.}(2018){Kowalska-Leszczynska},
  {Bzowski}, {Sok{\'o}{\l}}, \& {Kubiak}}]{kowalska-leszczynska_etal:18b}
{Kowalska-Leszczynska}, I., {Bzowski}, M., {Sok{\'o}{\l}}, J.~M., \& {Kubiak},
  M.~A. 2018, \apj, 868, 49, \dodoi{10.3847/1538-4357/aae70b}

\bibitem[{{Kubiak} {et~al.}(2014){Kubiak}, {Bzowski}, {Sok{\'o}{\l}},
  {Swaczyna}, {Grzedzielski}, {Alexashov}, {Izmodenov}, {Moebius}, {Leonard},
  {Fuselier}, {Wurz}, \& {McComas}}]{kubiak_etal:14a}
{Kubiak}, M.~A., {Bzowski}, M., {Sok{\'o}{\l}}, J.~M., {et~al.} 2014, \apjs,
  213, 29, \dodoi{10.1088/0067-0049/212/2/29}

\bibitem[{Kubiak {et~al.}(2016)Kubiak, Swaczyna, Bzowski, Sok{\'o}{\l},
  Fuselier, Galli, Heirtzler, Kucharek, Leonard, McComas, Park, Schwadron, \&
  Wurz}]{kubiak_etal:16a}
Kubiak, M.~A., Swaczyna, P., Bzowski, M., {et~al.} 2016, \apjs, 223, 35,
  \dodoi{10.1088/0067-0049/220/2/35}

\bibitem[{{Lallement} {et~al.}(2005){Lallement}, {Qu{\' e}merais}, {Bertaux},
  {Ferron}, {Koutroumpa}, \& {Pellinen}}]{lallement_etal:05a}
{Lallement}, R., {Qu{\' e}merais}, E., {Bertaux}, J.~L., {et~al.} 2005,
  Science, 307, 1447

\bibitem[{{Lee} {et~al.}(2009){Lee}, {Fahr}, {Kucharek}, {M{\"o}bius},
  {Prested}, {Schwadron}, \& {Wu}}]{lee_etal:09a}
{Lee}, M.~A., {Fahr}, H.~J., {Kucharek}, H., {et~al.} 2009, \ssr, 146, 275,
  \dodoi{10.1007/s11214-009-9522-9}

\bibitem[{{McCammon} {et~al.}(1983){McCammon}, {Burrows}, {Sanders}, \&
  {Kraushaar}}]{mccammon_etal:83a}
{McCammon}, D., {Burrows}, D.~N., {Sanders}, W.~T., \& {Kraushaar}, W.~L. 1983,
  \apj, 269, 107, \dodoi{10.1086/161024}

\bibitem[{{McComas} {et~al.}(2015{\natexlab{a}}){McComas}, {Bzowski}, {Frisch},
  {Galli}, {Izmodenov}, {Katushkina}, {Kubiak}, {Lee}, {Leonard}, {M{\"o}bius},
  {Park}, {Schwadron}, {Sok{\'o}{\l}}, {Swaczyna}, {Wood}, \&
  {Wurz}}]{mccomas_etal:15b}
{McComas}, D., {Bzowski}, M.~{Fuselier}, S., {Frisch}, P., {et~al.}
  2015{\natexlab{a}}, \apjs, 220, 22, \dodoi{10.1088/0067-0049/220/2/22}

\bibitem[{{McComas} {et~al.}(2015{\natexlab{b}}){McComas}, {Bzowski}, {Frisch},
  {Fuselier}, {Kubiak}, {Kucharek}, {Leonard}, {M{\"o}bius}, {Schwadron},
  {Sok{\'o}{\l}}, {Swaczyna}, \& {Witte}}]{mccomas_etal:15a}
{McComas}, D., {Bzowski}, M., {Frisch}, P., {et~al.} 2015{\natexlab{b}}, \apj,
  801, 28, \dodoi{10.1088/0004-637X/801/1/28}

\bibitem[{{McComas} {et~al.}(2013){McComas}, Angold, Elliott, Livadiotis,
  Schwadron, Skoug, \& Smith}]{mccomas_etal:13b}
{McComas}, D.~J., Angold, N., Elliott, H.~A., {et~al.} 2013, \apj, 779, 2

\bibitem[{{McComas} {et~al.}(2008){McComas}, Ebert, Elliot, Goldstein, Gosling,
  Schwadron, \& Skoug}]{mccomas_etal:08a}
{McComas}, D.~J., Ebert, R.~W., Elliot, H.~A., {et~al.} 2008, \grl, 35, L18103,
  \dodoi{10.1029/2008GL034896}

\bibitem[{{McComas} {et~al.}(2009){McComas}, {Allegrini}, {Bochsler},
  {Bzowski}, {Collier}, {Fahr}, {Fichtner}, {Frisch}, {Funsten}, {Fuselier},
  {Gloeckler}, {Gruntman}, {Izmodenov}, {Knappenberger}, {Lee}, {Livi},
  {Mitchell}, {M{\"o}bius}, {Moore}, {Pope}, {Reisenfeld}, {Roelof},
  {Scherrer}, {Schwadron}, {Tyler}, {Wieser}, {Witte}, {Wurz}, \&
  {Zank}}]{mccomas_etal:09a}
{McComas}, D.~J., {Allegrini}, F., {Bochsler}, P., {et~al.} 2009, \ssr, 146,
  11, \dodoi{10.1007/s11214-009-9499-4}

\bibitem[{{M{\"o}bius} {et~al.}(2004){M{\"o}bius}, {Bzowski}, {Chalov}, {Fahr},
  {Gloeckler}, {Izmodenov}, {Kallenbach}, {Lallement}, {McMullin}, {Noda},
  {Oka}, {Pauluhn}, {Raymond}, {Ruci{\'n}ski}, {Skoug}, {Terasawa}, {Thompson},
  {Vallerga}, {von Steiger}, \& {Witte}}]{mobius_etal:04a}
{M{\"o}bius}, E., {Bzowski}, M., {Chalov}, S., {et~al.} 2004, \aap, 426, 897,
  \dodoi{10.1051/0004-6361:20035834}

\bibitem[{{M{\"o}bius} {et~al.}(2009{\natexlab{a}}){M{\"o}bius}, {Kucharek},
  {Clark}, {O'Neill}, {Petersen}, {Bzowski}, {Saul}, {Wurz}, {Fuselier},
  {Izmodenov}, {McComas}, {M{\"u}ller}, \& {Alexashov}}]{mobius_etal:09a}
{M{\"o}bius}, E., {Kucharek}, H., {Clark}, G., {et~al.} 2009{\natexlab{a}},
  \ssr, 146, 149, \dodoi{10.1007/s11214-009-9498-5}

\bibitem[{{M{\"o}bius} {et~al.}(2009{\natexlab{b}}){M{\"o}bius}, {Bochsler},
  {Bzowski}, {Crew}, {Funsten}, {Fuselier}, {Ghielmetti}, {Heirtzler},
  {Izmodenov}, {Kubiak}, {Kucharek}, {Lee}, {Leonard}, {McComas}, {Petersen},
  {Saul}, {Scheer}, {Schwadron}, {Witte}, \& {Wurz}}]{mobius_etal:09b}
{M{\"o}bius}, E., {Bochsler}, P., {Bzowski}, M., {et~al.} 2009{\natexlab{b}},
  Science, 326, 969, \dodoi{10.1126/science.1180971}

\bibitem[{M{\"o}bius {et~al.}(2012)M{\"o}bius, Bochsler, Heirtzler, Kucharek,
  Lee, Leonard, Petersen, Schwadron, Valocvin, Wu, Bzowski, Kubiak, Fuselier,
  Saul, Wurz, McComas, \& Crew}]{mobius_etal:12a}
M{\"o}bius, E., Bochsler, P., Heirtzler, D., {et~al.} 2012, \apjs, 198, 11,
  \dodoi{10.1088/0067-0049/198/2/11}

\bibitem[{{M{\"o}bius} {et~al.}(2015){M{\"o}bius}, {Bzowski}, {Fuselier},
  {Heirtzler}, {Kubiak}, {Kucharek}, {Lee}, {Leonard}, {McComas}, {Schwadron},
  {Sok\'{o}{\l}}, \& {Wurz}}]{mobius_etal:15b}
{M{\"o}bius}, E., {Bzowski}, M., {Fuselier}, S.~A., {et~al.} 2015, \apjs, 220,
  24, \dodoi{10.1088/0067-0049/220/2/24}

\bibitem[{{Pogorelov} {et~al.}(2009){Pogorelov}, {Heerikhuisen}, {Mitchell},
  {Cairns}, \& {Zank}}]{pogorelov_etal:09a}
{Pogorelov}, N.~V., {Heerikhuisen}, J., {Mitchell}, J.~J., {Cairns}, I.~H., \&
  {Zank}, G.~P. 2009, \apjl, 695, L31, \dodoi{10.1088/0004-637X/695/1/L31}

\bibitem[{{Pogorelov} {et~al.}(2008){Pogorelov}, {Heerikhuisen}, \&
  {Zank}}]{pogorelov_etal:08a}
{Pogorelov}, N.~V., {Heerikhuisen}, J., \& {Zank}, G.~P. 2008, \apjl, 675, L41,
  \dodoi{10.1086/529547}

\bibitem[{{Redfield} \& {Linsky}(2008)}]{redfield_linsky:08a}
{Redfield}, S., \& {Linsky}, J.~L. 2008, \apj, 673, 283, \dodoi{10.1086/524002}

\bibitem[{{Richardson}(2008)}]{richardson:08a}
{Richardson}, J.~D. 2008, \grl, 35, 23104, \dodoi{10.1029/2008GL036168}

\bibitem[{Richardson {et~al.}(2008)Richardson, Liu, Wang, \&
  McComas}]{richardson_etal:08a}
Richardson, J.~D., Liu, Y., Wang, C., \& McComas, D. 2008, \aap, 491, 1,
  \dodoi{10.1051/0004-6361:20078565}

\bibitem[{{Schwadron} {et~al.}(2015){Schwadron}, {M{\"o}bius}, {Leonard},
  {Fuselier}, {Bzowski}, {Frisch}, {Heirtzler}, {Kubiak}, {Kucharek}, {Lee},
  {McComas}, {Rahmanifard}, {Sok\'{o}{\l}}, \& {Swaczyna}}]{schwadron_etal:15a}
{Schwadron}, N., {M{\"o}bius}, E., {Leonard}, T., {et~al.} 2015, \apjs, 220,
  25, \dodoi{0.1088/0067-0049/220/2/25}

\bibitem[{{Schwadron} {et~al.}(2016){Schwadron}, {M{\"o}bius}, {McComas},
  {Bochsler}, {Bzowski}, {Fuselier}, {Livadiotis}, {Frisch}, {M{\"u}ller},
  {Heirtzler}, {Kucharek}, \& {Lee}}]{schwadron_etal:16a}
{Schwadron}, N.~A., {M{\"o}bius}, E., {McComas}, D.~J., {et~al.} 2016, \apj,
  828, 81, \dodoi{10.3847/0004-637X/828/2/81}

\bibitem[{{Slavin} \& {Frisch}(2008)}]{slavin_frisch:08a}
{Slavin}, J.~D., \& {Frisch}, P.~C. 2008, \aap, 491, 53,
  \dodoi{10.1051/0004-6361:20078101}

\bibitem[{{Smith} {et~al.}(2014){Smith}, {Foster}, {Edgar}, \&
  {Brickhouse}}]{smith_etal:14a}
{Smith}, R.~K., {Foster}, A.~R., {Edgar}, R.~J., \& {Brickhouse}, N.~S. 2014,
  \apj, 787, 77, \dodoi{10.1088/0004-637X/787/1/77}

\bibitem[{{Snowden}(2015)}]{snowden:15a}
{Snowden}, S.~L. 2015, in Journal of Physics Conference Series, Vol. 577, J.
  Phys. Conf. Ser., 012022

\bibitem[{{Snowden} {et~al.}(1997){Snowden}, {Egger}, {Freyberg}, {McCammon},
  {Plucinsky}, {Sanders}, {Schmitt}, {Tr{\"u}mper}, \&
  {Voges}}]{snowden_etal:97a}
{Snowden}, S.~L., {Egger}, R., {Freyberg}, M.~J., {et~al.} 1997, \apj, 485,
  125, \dodoi{10.1086/304399}

\bibitem[{{Sok{\'o}{\l}} \& {Bzowski}(2014)}]{sokol_bzowski:14a}
{Sok{\'o}{\l}}, J.~M., \& {Bzowski}, M. 2014, ArXiv e-prints.
\newblock \doarXiv{1411.4826}

\bibitem[{{Sok{\'o}{\l}} {et~al.}(2019){Sok{\'o}{\l}}, {Bzowski}, \&
  {Tokumaru}}]{sokol_etal:19a}
{Sok{\'o}{\l}}, J.~M., {Bzowski}, M., \& {Tokumaru}, M. 2019, \apj, 872, 57,
  \dodoi{10.3847/1538-4357/aaf737}

\bibitem[{{Sok\'{o}{\l}} {et~al.}(2015{\natexlab{a}}){Sok\'{o}{\l}}, {Kubiak},
  {Bzowski}, \& {Swaczyna}}]{sokol_etal:15b}
{Sok\'{o}{\l}}, J.~M., {Kubiak}, M., {Bzowski}, M., \& {Swaczyna}, P.
  2015{\natexlab{a}}, \apjs, 220, 27, \dodoi{10.1088/0067-0049/220/2/27}

\bibitem[{{Sok\'{o}{\l}} {et~al.}(2015{\natexlab{b}}){Sok\'{o}{\l}}, {Bzowski},
  {Kubiak}, {Swaczyna}, {Galli}, {Wurz}, {M{\"o}bius}, {Kucharek}, {Fuselier},
  \& {McComas}}]{sokol_etal:15a}
{Sok\'{o}{\l}}, J.~M., {Bzowski}, M., {Kubiak}, M., {et~al.}
  2015{\natexlab{b}}, \apjs, 220, 29, \dodoi{10.1088/0067-0049/220/2/29}

\bibitem[{{Stone} {et~al.}(2013){Stone}, {Cummings}, {McDonald}, {Heikkila},
  {Lal}, \& {Webber}}]{stone_etal:13a}
{Stone}, E.~C., {Cummings}, A.~C., {McDonald}, F.~B., {et~al.} 2013, Science,
  341, 150, \dodoi{10.1126/science.1236408}

\bibitem[{Swaczyna {et~al.}(2019)Swaczyna, McComas, \&
  Schwadron}]{swaczyna_etal:19a}
Swaczyna, P., McComas, D., \& Schwadron, N. 2019, \apj, 871, 254,
  \dodoi{10.3847/1538-4357/aafa78}

\bibitem[{{Swaczyna} {et~al.}(2015){Swaczyna}, {Bzowski}, {Kubiak},
  {Sok\'{o}{\l}}, {M{\"o}bius}, {Leonard}, {Heirtzler}, {Kucharek},
  {Schwadron}, {Fuselier}, \& {McComas}}]{swaczyna_etal:15a}
{Swaczyna}, P., {Bzowski}, M., {Kubiak}, M., {et~al.} 2015, \apjs, 220, 26,
  \dodoi{10.1088/0067-0049/220/2/26}

\bibitem[{Swaczyna {et~al.}(2018)Swaczyna, Bzowski, Kubiak, Sok{\'o}{\l},
  Fuselier, Galli, Heirtzler, Kucharek, McComas, M{\"o}bius, Schwadron, \&
  Wurz}]{swaczyna_etal:18a}
Swaczyna, P., Bzowski, M., Kubiak, M., {et~al.} 2018, \apj, 854, 119,
  \dodoi{10.3847/1538-4357/aaabbf}

\bibitem[{{Tokumaru} {et~al.}(2015){Tokumaru}, {Fujiki}, \&
  {Iju}}]{tokumaru_etal:15a}
{Tokumaru}, M., {Fujiki}, K., \& {Iju}, T. 2015, \jgr, 120, 3283,
  \dodoi{10.1002/2014JA020765}

\bibitem[{{Vallerga} {et~al.}(2004){Vallerga}, {Lallement}, {Lemoine},
  {Dalaudier}, \& {McMullin}}]{vallerga_etal:04a}
{Vallerga}, J., {Lallement}, R., {Lemoine}, M., {Dalaudier}, F., \& {McMullin},
  D. 2004, \aap, 426, 855, \dodoi{10.1051/0004-6361:20035887}

\bibitem[{{Wilson} \& {Rood}(1994)}]{wilson_rood:94a}
{Wilson}, T.~L., \& {Rood}, R. 1994, \araa, 32, 191,
  \dodoi{10.1146/annurev.aa.32.090194.001203}

\bibitem[{{Witte}(2004)}]{witte:04}
{Witte}, M. 2004, \aap, 426, 835, \dodoi{10.1016/j.asr.2003.01.037}

\bibitem[{{Wolff} {et~al.}(1999){Wolff}, {Koester}, \&
  {Lallement}}]{wolff_etal:99}
{Wolff}, B., {Koester}, D., \& {Lallement}, R. 1999, \aap, 346, 969

\bibitem[{{Wood} {et~al.}(2015){Wood}, {M{\"u}ller}, \&
  {Witte}}]{wood_etal:15a}
{Wood}, B.~E., {M{\"u}ller}, H.-R., \& {Witte}, M. 2015, \apj, 801, 62,
  \dodoi{10.1088/0004-637X/801/1/62}

\bibitem[{{Zirnstein} {et~al.}(2016){Zirnstein}, {Heerikhuisen}, {Funsten},
  {Livadiotis}, {McComas}, \& {Pogorelov}}]{zirnstein_etal:16b}
{Zirnstein}, E.~J., {Heerikhuisen}, J., {Funsten}, H.~O., {et~al.} 2016, \apjl,
  818, L18, \dodoi{10.3847/2041-8205/818/1/L18}

\end{thebibliography}

\end{document}